\documentclass[preprint,5p,11pt]{elsarticle}

\usepackage{amsmath, amssymb, amsfonts, bm}
\usepackage{graphicx}
\usepackage{hyperref, cleveref}

\usepackage[dvipsnames]{xcolor}
\usepackage{kotex,soul}
\setstcolor{red}

\journal{Chaos, Solitons and Fractals}
\begin{document}

\begin{frontmatter}
\title{Hybrid Synchronization with Continuous Varying Exponent in Decentralized Power Grid}

\author[1]{Jinha Park}
\author[1]{B. Kahng\corref{cor1}}
\ead{bkahng@kentech.ac.kr}
\cortext[cor1]{corresponding author}
\affiliation[1]{
    organization={CCSS, KI for Grid Modernization, Korea Institute of Energy Technology},
    city={Naju},
    postcode={58217},
    state={Jeonnam},
    country={Korea}
}

\begin{abstract}
Motivated by the decentralized power grid, we consider a synchronization transition (ST) of the Kuramoto model (KM) with a mixture of first- and second-order type oscillators with fractions $p$ and $1-p$, respectively.
Discontinuous ST with forward-backward hysteresis is found in the mean-field limit. A critical exponent $\beta$ is noticed in the spinodal drop of the order parameter curve at the backward ST. We find critical damping inertia $m_*(p)$ of the oscillator mixture, where the system undergoes a characteristic change from overdamped to underdamped. When underdamped, the hysteretic area also becomes multistable. This contrasts an overdamped system, which is bistable at hysteresis. We also notice that $\beta(p)$ continuously varies with $p$ along the critical damping line $m_*(p)$. Further, we find a single-cluster to multi-cluster phase transition at $m_{**}(p)$. We also discuss the effect of those features on the stability of the power grid, which is increasingly threatened as more electric power is produced from inertia-free generators.
\end{abstract}



\end{frontmatter}

The Kuramoto model (KM) has served as a paradigmatic model of synchronization phenomena~\cite{Kuramoto}, ranging from biological to social systems~\cite{Winfree,Strogatz,Vicsek,Ermentrout,pedestrian}, and contributed significantly to our understanding of the nature of synchronization transition (ST). Many studies focus on the KM without an inertia term. There, the oscillators undergo overdamped motions. Continuous~\cite{Kuramoto,Winfree,Strogatz,Vicsek,Ermentrout,pedestrian}, discontinuous~\cite{martens,bellerophon}, or hybrid ST~\cite{Pazo,song,metastable,physicaD,leader} occurs, depending on the types of intrinsic frequency distribution of oscillators. On the other hand, KM with inertia is known for a discontinuous ST~\cite{Tanaka1,Tanaka2,Gao1,Olmi,boccaletti,Gao2}. A nonzero inertia introduces angular momentum, a tendency to sustain its oscillatory motion, represented by a second-order time derivative term in the equation of motion. In this regard, we call the oscillators with (without) inertia second-order (first-order). Likewise, a first-order (second-order) KM comprises first-order (second-order) oscillators.

In the pure first-order system, the intrinsic frequency distribution determines the type of ST. Notably, each mode of a multi-modal frequency distribution of a first-order KM can seed competition. An unimodal distribution forms a single center of gathering at the mode, and the ST is triggered and led by the oscillators with natural frequencies closer to the mode. A single giant cluster emerges from this seed, growing continuously without competition by merging the oscillators with frequencies nearby as the strength of the interaction is increased. In contrast, a bimodal system organizes a competing pair of giant oscillators~\cite{martens}.
Consequently, multiple clusters characterized by different average frequencies can coexist before merging into one giant cluster. A related phenomenon is standing wave~\cite{martens} or bellerophon states~\cite{bellerophon}. The order parameter curve shows a discontinuous jump with hysteresis. A uniform distribution can be considered an infinite collection of equally spaced modes. Each oscillator becomes a competing seed that tries to trigger the synchronization. With a uniform frequency distribution, a unanimous locking-unlocking occurs suddenly at the hybrid ST point~\cite{Pazo}. The dynamic pathway that joins the two phases can be associated with crossing a flat \textit{ad hoc} potential, making the system highly susceptible and critical~\cite{song}.

On a network, a hub node with a higher degree of connection is more likely to become a seed for synchronization because it interacts more with the system. The synchronization cluster grows and extends from the center to the periphery~\cite{yamir}, generating a continuous ST. Discontinuous ST occurs if a finite fraction of oscillators in the system suddenly lock all together at the transition point. Structural suppression is to set the hub frequencies relatively higher so that the barrier to synchronization is increased and offsets its geometric tendency to synchronize better. If the frequencies of the nodes on an oscillator network are adjusted proportionately to their degrees, the synchronizability becomes more equalized. Such an adjustment or similar counteractions can delay the onset of synchronization and potentially lead to an explosive ST~\cite{dfckm,Ji,zhang,zhang2,zhang3}.
Further, a hybrid ST may occur as a special case. For example, with a particular choice of the degree exponent, $\lambda=3$ for the degree-frequency correlated scale-free KM, the mean angular velocities of the oscillators are uniformly equalized~\cite{dfckm,gaoHPT}. Consequentially, the model maps to the Paz\'{o} case with uniform frequency distribution~\cite{dfckm,gaoHPT}, which is known for the hybrid ST with $\beta=2/3$~\cite{Pazo}.

A second-order KM, on the other hand, undergoes a discontinuous ST~\cite{Tanaka1,Tanaka2,Gao1,Olmi,boccaletti,Gao2}. Hysteresis of pinning and depinning can cause differences in the forward and backward transition points~\cite{Tanaka2}. Microscopically, a single underdamped second-order oscillator can become bistable~\cite{Strogatz,Goldstein,Belykh}. A collection of many bistable oscillators generates rich synchronization dynamics bearing multistability, and the steady state strongly depends on the protocol~\cite{Olmi}. Moreover, under extreme damping, the single-frequency cluster state becomes unstable for high inertia systems. The synchronization cluster split into multiple clusters, and the coherence of the system becomes nonstationary, showing large-amplitude oscillations in $R(t)$ even in the steady state~\cite{Tanaka1,Olmi}, as the clusters revolve around each other.

We claim that mixing two oscillator types can potentially generate a discontinuous ST or a hybrid ST~\cite{metastable, physicaD,leader}. One type promotes or leads the ST, while the other type relatively suppresses or delays the ST. A tug-of-war between the two entities may include critical phenomena at the transition point. Moreover, cooperative effects of the two types can generate new orders, such as traveling wave~\cite{metastable}. In this paper, we investigate a Kuramoto oscillator mixture, where one type has inertia, and the other has no inertia. Depending on the presence and magnitude of inertia, each oscillator alters its dynamical character: overdamped or underdamped. An overdamped oscillator quickly adapts its dynamical state to the external driving of coupling strength. An underdamped oscillator, on the contrary, may follow behind the schedule. If the two types are mixed, overdamped oscillators promote synchronization, while underdamped oscillators suppress the ST. The consequence of a competition between the two tendencies will depend on the mixture compositions. In this regard, we focus on the tug-of-war of Kuramoto oscillator mixtures, where one type has inertia, and the other has no inertia. Depending on the mixed fraction of the oscillators with different characters, the ST's transition point and jump size are altered. Moreover, we determine whether the mixture oscillator system is overdamped or underdamped. Finally, we investigate the possibility of a hybrid ST for a mixture KM, a mixture of first-order KM and second-order KM, which are known to exhibit continuous ST and discontinuous ST, respectively.

This paper considers a mixed-order KM composed of $p$-fraction of oscillators without inertia and $(1-p)$-fraction of oscillators with inertia. For simplicity, we assume that every second-order oscillator has an equal inertia $m$. Also, we consider a fully connected oscillator system, which excludes the structural heterogeneities and enables us to focus solely on the suppressive effects of inertia. It is remarked that in general, on a network, however, ST may delicately depend on how the inertia is distributed~\cite{Laurent,Motter} and stability and multistability issues become more intricate subjects~\cite{Motter,powergrid,powergrid2,braess,basin,survivability}. Overdamped oscillators promote ST, while underdamped oscillators suppress ST. As the number of suppressive population $1-p$ grows, the inertial suppression delays the transition point $K_c$ and increases the jump size $R_c$.

\begin{figure}
\centering
\includegraphics[width=0.6\linewidth]{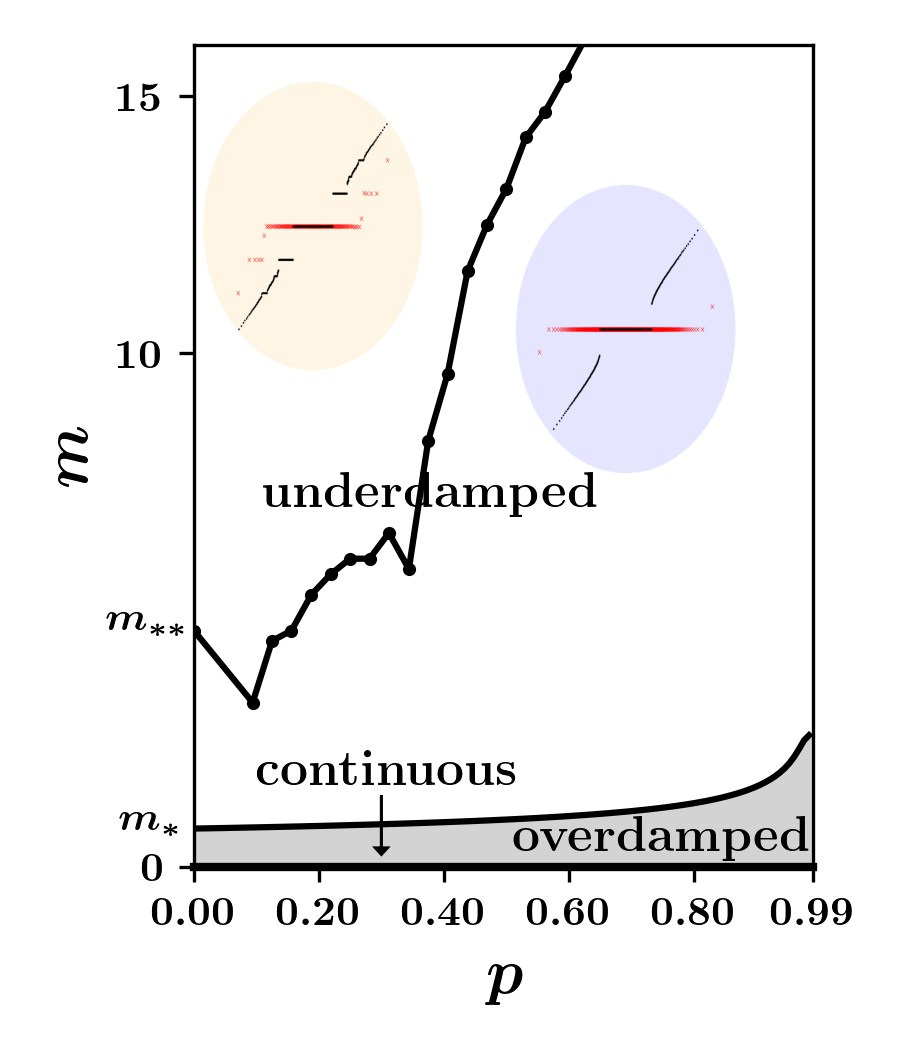}
\caption{
    Phase diagram of the mixed-order KM, with parameters $(p,m)$. The critical curve dividing the overdamped and underdamped regions is estimated as $m_*(p)\approx 0.74 (1-p)^{-0.32}$. Across the curve, the bistabilities of the second-order oscillators emerge or vanish. At $m=0$ or $p=1$, the mixture system is reduced to a pure first-order KM and exhibits a continuous ST. At $m_{**}(p)$, there is a transition from a coherent phase with a single frequency cluster to another coherent phase with multiple frequency clusters. The thumbnails are $\langle\dot\theta\rangle$ versus $\omega$, which reveals typical multi- and single-frequency clustering patterns. The red (black) symbols correspond to the first (second)-order oscillators. Notice that in the single-cluster pattern in the underdamped region, first-order oscillators (red) have a larger window of frequency entrainment compared to the second-order oscillators (black). It is remarked that the single cluster pattern also occurs in the synchronization phase, but the widths of entrainment become the same. The boundary $m_{**}(p)$ is obtained from a single configuration. See Fig.~\ref{fig:multiclusters} in the Supplementary Information for how $m_{**}$ were obtained.
}   \label{fig:critical-line}
\end{figure}

The following results are obtained: the overall phase transition is discontinuous and hysteretic unless the model trivially reduces to a first-order model ($m=0$ or $p=1$). ST is also path (protocol)-dependent. (i) the forward transition is discontinuous, (ii) the backward transition is hybrid. The system is overdamped, critically damped, underdamped, or extreme-underdamped, depending on the model parameters $(m,p)$. (iii) the overdamped system has a unique, coherent state, (iv) the underdamped system has a range of multistable stationary coherent states between the forward and backward curves, and its steady state strongly depends on the initial conditions. (v) The extreme underdamped system exhibits a nonstationary coherent state. (vi) Along the curve of critical damping, we find a continuously varying backward hybrid critical exponent $\beta$ depending on the fraction $p$. The transition lines between overdamped, underdamped, and multi-cluster phases are shown in Fig.~\ref{fig:critical-line}. We calculate and numerically find two characteristic values of inertia $m_*$ and $m_{**}$. Forward-backward hysteresis and multistability are found in the underdamped region $m_{*}< m < m_{**}$. The critical damping $m_*(p)$ and cluster splitting $m_{**}(p)$ are $p$-dependent.

The mixed-order KM is written as follows:
\begin{align}
    \mu_i\ddot \theta_i + \gamma \dot\theta_i = \omega_i + \frac{K}{N}\sum_{j=1}^{N}\sin(\theta_j-\theta_i),
    \label{eq:model}
\end{align}
where $\mu_i$ is the rotational inertia of oscillator $i$ ($i=1,2,\cdots,N$), $\gamma$ is dissipation strength, $\omega_i$ is the intrinsic frequency, and $K$ is the coupling strength. $\mu$ and $\omega$ are sampled from the distribution
\begin{align}
    g(\mu,\omega) = h(\omega)\big[p\delta(\mu) + (1-p) \delta(\mu-m)\big],
\end{align}
where $p\equiv N_1/N$ is the fraction of first-order type oscillators without inertia, and $1-p\equiv N_2/N$ is the fraction of second-order type oscillators with homogeneous values of inertia $m$, and $h(\omega)=\exp(-\omega^2/2)/\sqrt{2\pi}$ is the standard Gaussian distribution. The damping $\gamma$ is set to unity without loss of generality. The synchronization order parameters $R$ and $R_a$ are defined as
\begin{align}
    Re^{i\psi}= \frac{1}{N}\sum_{j=1}^{N} e^{i\theta_j} \quad \textrm{and} \quad
    R_{a}e^{i\psi_a}= \frac{1}{N_a}\sum_{j}^{N_a}e^{i\theta_j},
\end{align}
for the first- and second-order oscillators $a=1,2$, respectively, and we set $\psi=\psi_{1}=\psi_2=0$ without loss of generality. Then, for each of the second-order oscillators $\mu_i=m$ in the rotating frame, Eq.~\eqref{eq:model} can be rewritten in a standard form
\begin{align}
    \frac{d^2\theta}{d\tau^2} + a \frac{d\theta}{d\tau} & = b - \sin \theta
\end{align}
with two effective parameters
\begin{align}
    a = \frac{\gamma}{\sqrt{mKR}} \quad \textrm{~and} \quad b = \frac{\omega-\gamma\Omega}{KR}
\end{align}
and rescaled time $\tau=t\sqrt{KR/m}$. $\Omega=\lim_{\tau\to\infty}\psi_2/\tau$ is zero because the mean frequency of the oscillators $\overline{\omega}$ is zero and $K$ is positive~\cite{metastable,contrarian}.

\begin{figure}
\centering
\includegraphics[width=\linewidth]{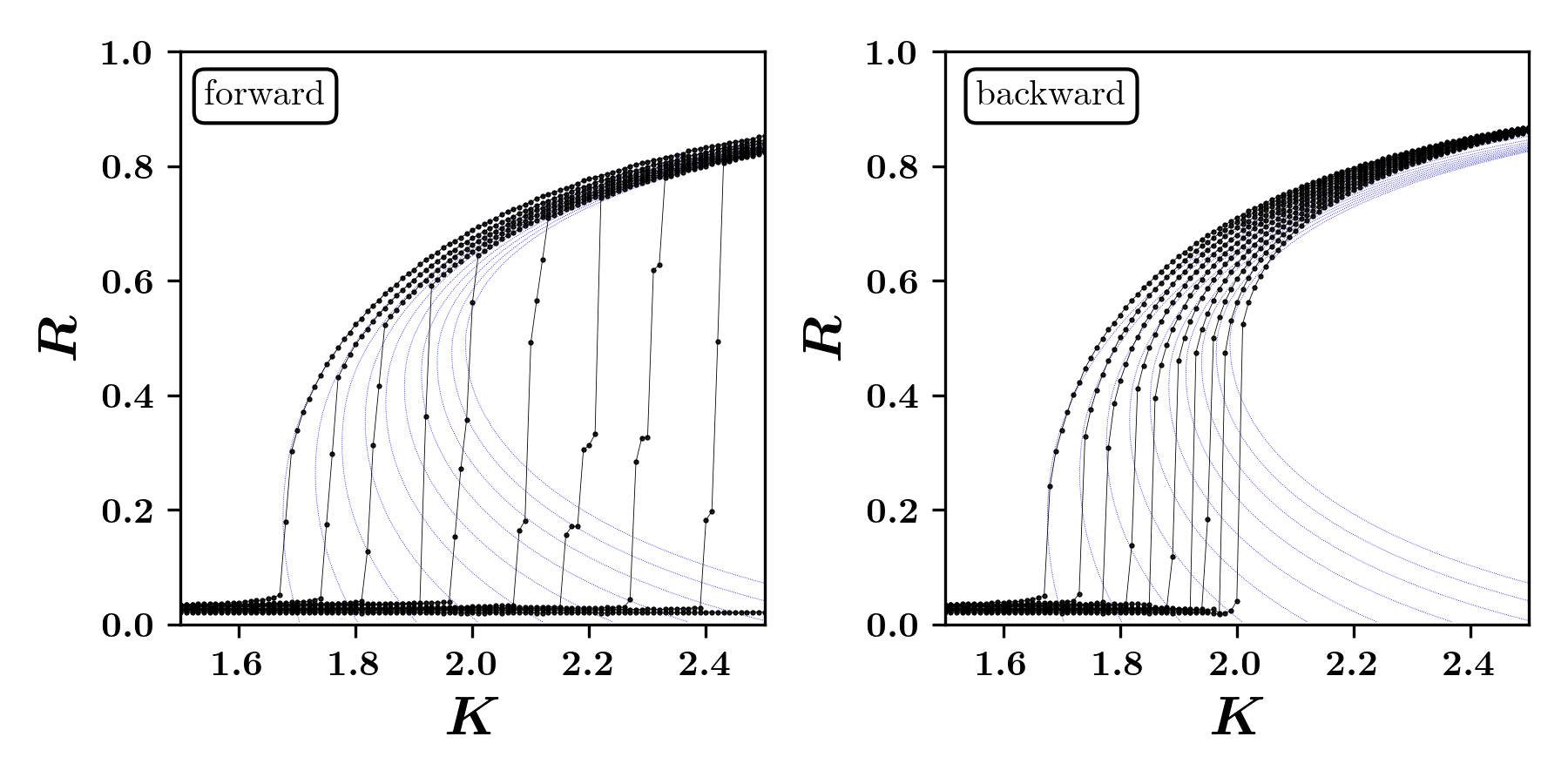}
\caption{
    Order parameter curves at the critical mass $m_*(p)$ for $p=0.0,0.1,\cdots,0.9$, from right to left. As suppressive fraction $1-p$ increases, the ST is delayed, and the jump size is increased. We obtain continuously varying $\beta(p)$ from the backward curves, and the values are summarized in Table~\ref{table:saddle-pt}. Black symbols denote (a) forward (b) backward Runge-Kutta simulation results. $N=1024$, averaged over $N_{ens}=5$ independent runs. Blue dotted curves are self-consistency solutions.
}   \label{fig:op-curvep}
\end{figure}

The steady states of Eq.~\eqref{eq:model} with stationary values of $R$ can be solved using the self-consistency method~\cite{Tanaka1,Tanaka2,Gao1}. The total coherence $R$ is the sum of the contribution $R_1$ of the first-order oscillators and the contribution $R_2$ of the second-order oscillators with their weights. In the thermodynamic limit $N\to\infty$, the self-consistency equation (SCE) of the synchronization order parameter $R$ is written as~\cite{Tanaka1,Tanaka2,Gao1}
\begin{align}   \label{eq:sce}
    R & = p\int_{-1}^{1}\tilde{h}(b) \sqrt{1-b^2}db \nonumber                                 \\
      & + (1-p)\int_{-b_{c}}^{b_{c}}\tilde{h}(b)  \sqrt{1-b^2}db \nonumber                    \\
      & - (1-p)\int_{-\infty}^{\infty}\tilde{h}(b) \Theta(b_c-|b|) \frac{a^4}{2(a^4+b^2)} db,
\end{align}
where $\tilde{h}(b)\equiv KR h(KRb)$ and $b_c\equiv \omega_c/KR$, and $\omega_c$ borders the locking or drifting of the bistable oscillators. Here we have made a strong assumption that $|\omega|$ in the range $[0,\omega_c]$ are locked, and $[\omega_c,KR]$ are drifting. It is also important to note that Eq.~\eqref{eq:sce} is still undetermined because the value of $\omega_c$ is unknown. Following Refs.~\cite{Tanaka1,Tanaka2,Gao1}, we further assume that $\omega_c$ is equal to the homoclinic bifurcation $\omega_h$ in the forward process and $\omega_c$ is equal to $KR$ in the backward process. Notice that the first and second terms of the SCE~\eqref{eq:sce} are the contributions of the locked oscillators. The third term is the contribution of the limit-cycling second-order oscillators~\cite{Gao1}. The latter contribution is relatively small, depends on $m$ in the orders of magnitude $\sim m/(1+m^2)$, and it gives a negative contribution to the total coherence. It is remarked that the drifting first-order oscillators $|\omega|>KR$, in contrast, do not contribute to the total coherence $R$. The system's total coherence $R(K,m,p)$ depends on the coupling strength $K$, rotational inertia $m$ of the second-order oscillators, and the fraction $p$ of first-order oscillators as shown in Fig.~\ref{fig:op-curvep}.

\begin{figure}
\centering
\includegraphics[width=\linewidth]{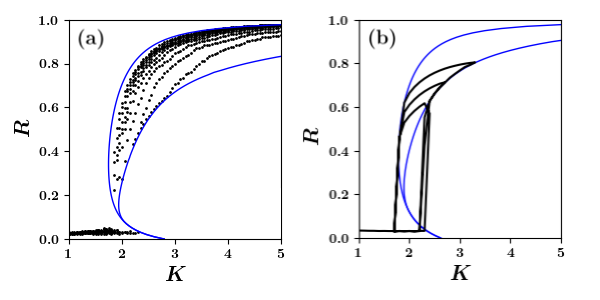}
\caption{
    (a) Coherent state of an underdamped system ($p=0.5,m=5$) may become multistable. Depending on the initial condition, a range of steady-state coherence value $R_f(K)\leq R \leq R_b(K)$ is possible for each $K$. $m=5, p=0.5$ (b) Hysteresis depends on the driving protocol in underdamped systems. Here, a protocol $\{R_0, R_1,\cdots\}=\{0, 0.8, 0.05, 0.7, 0.2, 0.5, 0.4\}$ is taken for an underdamped system $p=0.5,m=3$ starting at an initially random phase at $K=0$. The coupling constant is increased and decreased.
}   \label{fig:multistability}
\end{figure}

The percolation order parameter $G$ is the fraction of oscillators in the giant frequency cluster. Note that this locking fraction can be calculated as
\begin{align}
    G & = p\int_{-1}^{1}\tilde{h}(b) db + (1-p)\int_{-b_{c}}^{b_{c}}\tilde{h}(b),
    \label{eq:perc-op}
\end{align}
using the self-consistent $R$ of Eq.~\eqref{eq:sce}. We present an iterative algorithm to find a self-consistency solution efficiently on a multi-dimensional domain. The solutions are cross-checked with direct numerical integration of Eq.~\eqref{eq:model}. See Methods for more details. The forward and backward self-consistent order parameter curves are defined using $\omega_c=\omega_h$ and $\omega_c=KR$, respectively. The forward and backward self-consistent order parameter curves may either become separated, as shown in Fig.~\ref{fig:multistability}, or they may collapse into an identical curve.

\begin{figure}[t]
\centering
\includegraphics[width=\linewidth]{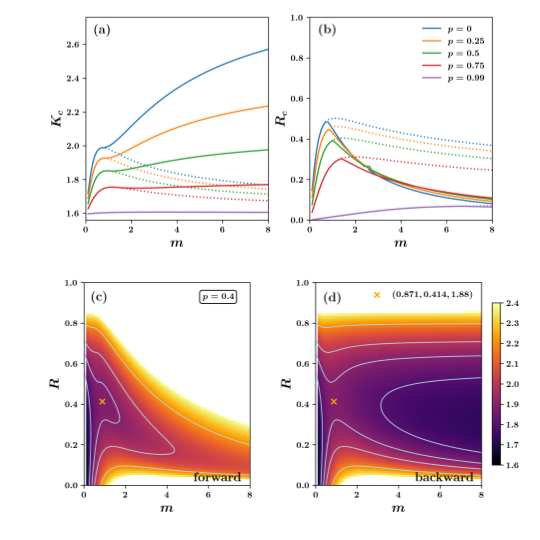}
\caption{
    The heatmap of the self-consistent solution $K(m,R)$ for $K$ in the range $[1.6,2.4]$ and $p=0.4$. Contours are drawn at $K=1.8,1.9,2.0,2.2$. Notice a saddle at $m_*=0.87$, marked with ($\times$). (a) Transition point and (b) jump size of the order parameter in the forward (solid) and backward (dotted) processes at specified values of $p$. Multistability disappears (emerges) at $m=m_*$. Panels (c) and (d) correspond to forward and backward processes, respectively.
}   \label{fig:mrK-heatmap}
\end{figure}

How can we define and calculate the critical inertia $m_*$ of the mixture system?
At the critical inertia $m_*$, the bistable phase of an individual second-order oscillator emerges. Recall that homoclinic, saddle-node, and infinite-period bifurcations meet at one point in the phase diagram (see Fig.1 of ref.~\cite{Gao1}), where the bistable phase ends. We notice that $a_*=\gamma/\sqrt{m_*KR}$ associates with the critical inertia $m_*$ of the second-order system, which divides the phase diagram into underdamped $a<a_*$ ($m>m_*$) and overdamped $a>a_*$ ($m<m_*$) regions.
Similarly, the mixture system we consider here is critically damped where the multistability emerges. This occurs where the forward and backward self-consistent order parameter curves $R_f(K)$ and $R_b(K)$ separate. Observe in Fig.~\ref{fig:mrK-heatmap} (a) and (b) that forward (solid) and backward (dotted) transition point $K_c$ and jump size $R_c$ is the same for $0\leq m \leq m_*$. The location of separation also corresponds to the location of the saddle of the iso-surface $K(m,R,p)$, as shown in Fig.~\ref{fig:mrK-heatmap} (c) and (d).

Now, the critical inertia of the mixed-order system is specified by a saddle. The critical inertia $m_*$ of the mixture becomes $p$-dependent. This definition of $m_*$ results in a consistent value of $a_*\simeq 1.193$ at $p=0$, in the pure second-order KM, obtained by Melnikov's method~\cite{Belykh,Gao1}. We use Eq.~\eqref{eq:sce} to find the contours $K(m,R,p)$ for given $(m,R,p)$, as shown in Fig.~\ref{fig:mrK-heatmap}. There, the saddle $(m_*,R_*,K_*)$ is marked with $\times$ for given $p=0.4$ in Fig.~\ref{fig:mrK-heatmap} (c) for forward and (d) backward processes. In Fig.~\ref{fig:mrK-heatmap} (d), the curve is split into the top and bottom of the saddle for $K>K_*$. The coupling strength is sufficiently large so that the coherence is guaranteed for all values of $m$. It is mentioned that the curves below the saddle are unstable solutions not realized in numerical simulations. When $K<K_*$, the solution is split into the left and right of the saddle. The point with $\partial m/\partial R =0$ on the curves left (right) of the saddle corresponds to the forward (backward) transition point $K_c^f$ ($K_c^b$) and jump size $R_c^f$ ($R_c^b$) at the corresponding value of $m$, as shown in the solid (dotted) curves in Fig.~\ref{fig:mrK-heatmap} (a) and (b). For each given mixture fraction $p$, the backward transition point $K_c$ is maximally delayed, and also, the jump size is maximal at the critical inertia $m_*$. Also, $K_c(m_*)$ and $R_c(m_*)$ increases as the inertial population $1-p$ increases. It is important to note that multistability is absent when $m\leq m_*$. The precise location of the saddle is obtained numerically by the stationary state of the gentlest ascent dynamics~\cite{GAD} defined on the surface $K(m,R,p)$. See Methods for more details. The critical inertia $m_*$ and $a_*$ increase with $p$ (see Table~\ref{table:saddle-pt}).
Based on the critical inertia of the mixture KM $M_*(m,p) = Nm_*(1-p)$, one can determine whether the oscillator system as a whole is overdamped or underdamped. It is remarked that not only the total inertia matters, but also how the inertia is distributed across the system can affect the overdamped or underdamped character of the oscillator system. For instance, we can compare the mixture system of $1-p$ to $p$ fraction with the homogeneous system of equal total inertia. A mixture system $(p,m)=(0.5,1)$ is underdamped. In contrast, the pure system $(1-p,m)=(1,0.5)$ is overdamped. We anticipate that the heterogeneous distribution of inertia across the network may further become a critical factor that affects the synchronization of a complex oscillator system~\cite{Laurent}.

\begin{figure}
\centering
\includegraphics[width=\linewidth]{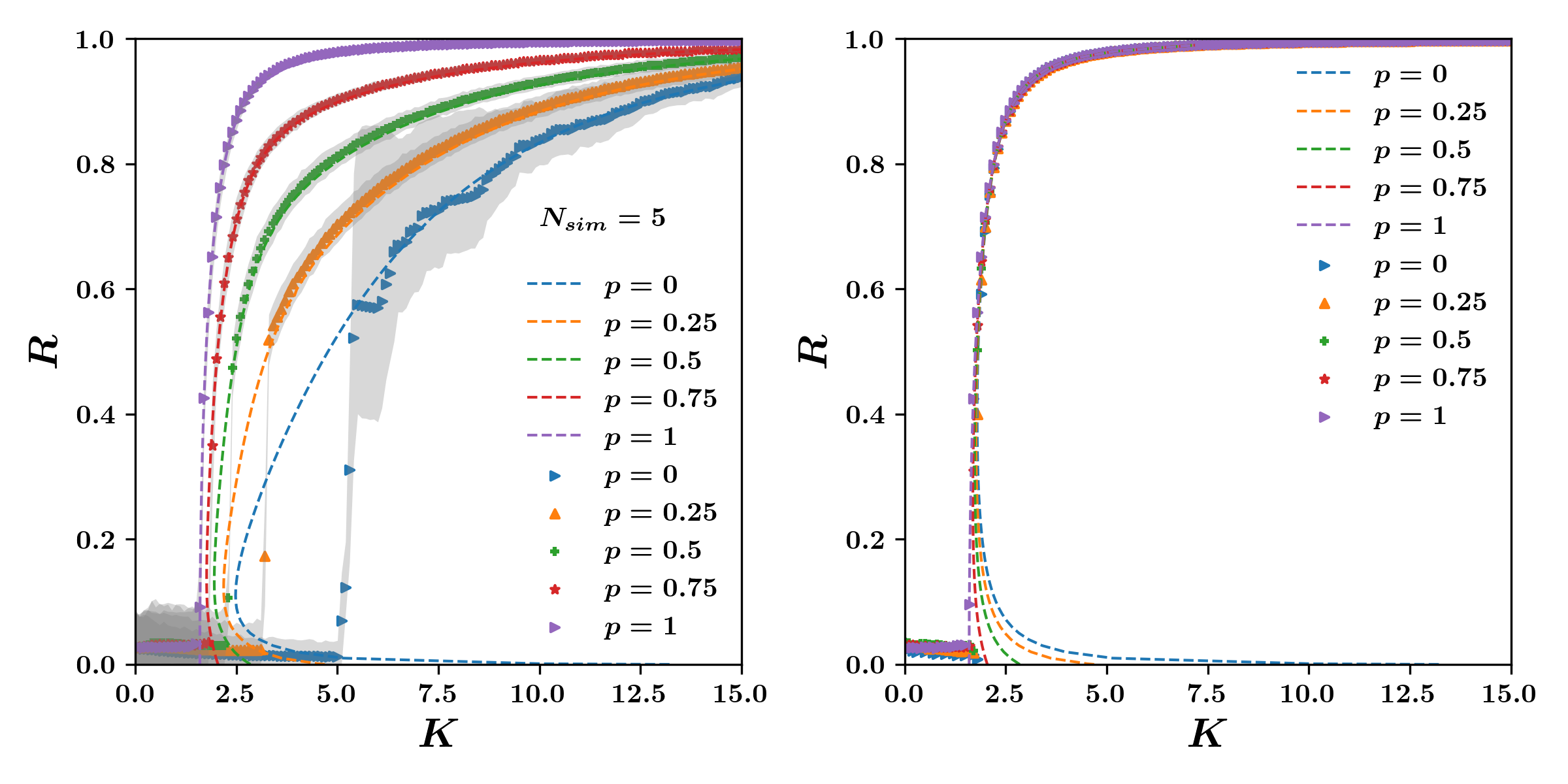}
\caption{
    Order parameter curves are shown for $m=6$ at specified values of $p$. Note that $m=6$ belongs to the underdamped region in Fig.~\ref{fig:critical-line}. Self-consistency solutions of Eq.~\eqref{eq:sce} are shown in dashed curves. The left and right panels correspond to forward and backward processes, respectively. Symbols result from simulations; the order parameter is time-averaged during the later half of the run time at each $K$ step and averaged over $N_{ens}=5$ runs with different initial random phases at $K=0$. $(N_1,N_2)=(0,1024)$,$(256,768)$,$(512,512)$,$(768,256)$,$(1024,0)$ and $\gamma=1$ were used. The gray shaded area on the left panel shows possible large fluctuations ($R_{min}$ to $R_{max}$) of the order parameter in the steady state, which relates to the birth of secondary clusters~\cite{Tanaka2,Olmi}. Secondary clusters dissipate, and the order parameter fluctuations become negligible when $p$ becomes larger. See~\ref{fig:clustering-underdamped-forward} for more details.
}   \label{fig:checking-sce-with-simulation}
\end{figure}

Having or not having multistability in the hysteresis is the key difference between the underdamped and overdamped systems. It is reiterated that forward and backward Runge-Kutta simulation curves are different in Fig.~\ref{fig:checking-sce-with-simulation}, even though forward and backward self-consistency solution curves coincide at each $p$. In Fig.~\ref{fig:multistability}(a), the coupling strength $K$ is fixed. However, depending on the initial conditions of the oscillators $\{\theta_i,\dot\theta_i\}_{i=1}^N$, the steady state order parameter $R$ takes a broad range of values bounded by the forward and backward self-consistent order parameter curves. In Fig.~\ref{fig:multistability}(b), one also observes the path-dependent memory effect (See Methods). After a return from multiple forward and backward drivings, the system's coherence can be altered from what it was before. The origin of such a path dependence is the bistability of the second-order oscillators. Bistability is a distinguishing feature of a second-order oscillator. A first-order oscillator has a steady state, stationary or drifting, determined solely by the ratio $b=\omega/KR$, regardless of the initial conditions. A second-order oscillator with bistability can take one of the two steady-state choices: becoming stationary at a fixed phase $\theta_{fp}$ or running a limit cycle $\dot\theta_{lc}(\theta)$~\cite{Strogatz,Goldstein}. The choice depends sensitively on the initial conditions. When driven through forward and backward processes, they collectively switch their steady states and result in multistability and complex path dependence in the system's steady state $R$, as noticed in Fig.~\ref{fig:multistability}(b) and also in Ref.~\cite{Olmi}. The microscopic difference in the forward and backward clustering patterns is noticed in Figs.~\ref{fig:clustering-underdamped-forward} and~\ref{fig:clustering-underdamped-backward} for an underdamped system. In contrast, locking and unlocking occur at the same transition point for an overdamped system. See Fig.~\ref{fig:clustering-overdamped}.

\begin{table}
    \centering
    \setlength{\tabcolsep}{2pt} 
    \begin{tabular}{ccccccc}
        \hline
        $p$  & $m_*$    & $R_*$    & $K_*$    & $a_*$    & $\beta_R$ & $\beta_G$ \\
        \hline
        0.0  & 0.746506 & 0.489413 & 1.988214 & 1.173314 & 0.44      & 0.44      \\
        0.1  & 0.770096 & 0.473848 & 1.964788 & 1.181000 & 0.45      & 0.44      \\
        0.2  & 0.797726 & 0.456351 & 1.939732 & 1.190016 & 0.45      & 0.44      \\
        0.3  & 0.830717 & 0.436466 & 1.912778 & 1.200786 & 0.45      & 0.44      \\
        0.4  & 0.871097 & 0.413559 & 1.883581 & 1.213964 & 0.45      & 0.44      \\
        0.5  & 0.922182 & 0.386708 & 1.851674 & 1.230602 & 0.45      & 0.45      \\
        0.6  & 0.989888 & 0.354496 & 1.816395 & 1.252552 & 0.45      & 0.45      \\
        0.7  & 1.086192 & 0.314548 & 1.776739 & 1.283487 & 0.46      & 0.45      \\
        0.8  & 1.240831 & 0.262311 & 1.731002 & 1.332254 & 0.46      & 0.45      \\
        0.9  & 1.564433 & 0.186517 & 1.675639 & 1.430120 & 0.46      & 0.46      \\
        0.99 & 2.445794 & 0.037392 & 1.606928 & 2.608568 & 0.47      & 0.47      \\
        \hline
    \end{tabular}
    \caption{
        Continuously varying critical exponents $\beta_R(p)$ and $\beta_G(p)$ along the critical curve $m=m_*(p)$. The critical inertia is calculated from the location of the saddle for different values of mixing fraction $0\leq p\leq 0.99$ as in Fig.~\ref{fig:mrK-heatmap}. $a_*$ is calculated from the definition $a=\gamma/\sqrt{mKR}$ where $\gamma$ is set to unity. The critical exponents $\beta_R$ and $\beta_G$ are calculated at the saddle for the order parameters $R$ and $G$, respectively, at the saddle for each $p$.
    }   \label{table:saddle-pt}
\end{table}

In the backward protocol, we observe a hybrid ST from a coherent state
\begin{align}
    R \sim R_c + c(K-K_c)^{\beta_R}, \quad \textrm{as } \quad K\to K_{c}^+,
    \label{eq:hpt}
\end{align}
to an incoherent state $R=0$. $R_c$ is the jump size in the synchronization order parameter, $c$ is a constant, and $\beta_R$ is the exponent of the order parameter curve at $K_c^+$ after the jump $R_c$. The exponent $\beta_R$ becomes critical if $\partial K/\partial R=0$ as $R\to R_c^+$. It should be strictly less than unity to be considered critical. A discontinuous transition with non-critical exponent $\beta_R=1$ occurs in the forward protocol. For the critical damping at $m=m_*(p)$, the hybrid critical exponent $\beta_R(p)$ continuously varies with dependence on $p$. Its value ranges from $\beta_R=0.44$ in the completely second-order system ($p=0$) to $\beta_R\to 1/2$ as the system becomes similar to a first-order system ($p=1$). See Fig.~\ref{fig:measurement-of-beta} for details. The critical exponent, $\beta_G$ for the percolation order parameter $G(K)$, can be defined similarly. The $p$ dependence of the critical inertia $m_*$, the corresponding jump size $R_*$, coupling strength $K_*$, and hybrid critical exponents $\beta_R$ and $\beta_G$ are summarized in Table~\ref{table:saddle-pt}.

It is remarked that a hybrid ST usually appears at a rare incidence, usually confined along the boundary of continuous ST and discontinuous ST regions, e.g., when the degree exponent of the scale-free network is exactly   $\lambda=3$~\cite{dfckm}, or when the curvature of frequency distribution becomes precisely flat~\cite{Pazo}. In contrast, our model shows a broad domain of hybrid ST without requiring a peculiar setting of the parameters; the critical exponent $\beta(p)$ is self-organized throughout the domain, varying continuously, depending on the suppression strength $1-p$. To our knowledge, Hybrid ST with continuously varying exponents is a novel phenomenon in synchronization contexts. It is remarked that a similar progression has been noticed in the context of hybrid percolation transition~\cite{rER,mBFW,IET}, where both forward~\cite{rER} and reverse~\cite{rrER} models turn out to exhibit hybrid phase transitions. It is also remarked that the hybrid critical exponent $\beta_R$ of the mixture model is comparatively smaller than that of the first-order model with uniform frequency distribution ($\beta=2/3$)~\cite{Pazo}. Thus, the coherence level is more sensitive to the change of $K$ near the transition point on the supercritical side. Massive movements are more disruptive. It is remarked that a hybrid ST occurs even for the unimodal natural frequency distribution in the mixed-order system, and thus, we anticipate that there lies a different microscopic mechanism.

\begin{figure}
\centering
\includegraphics[width=\linewidth]{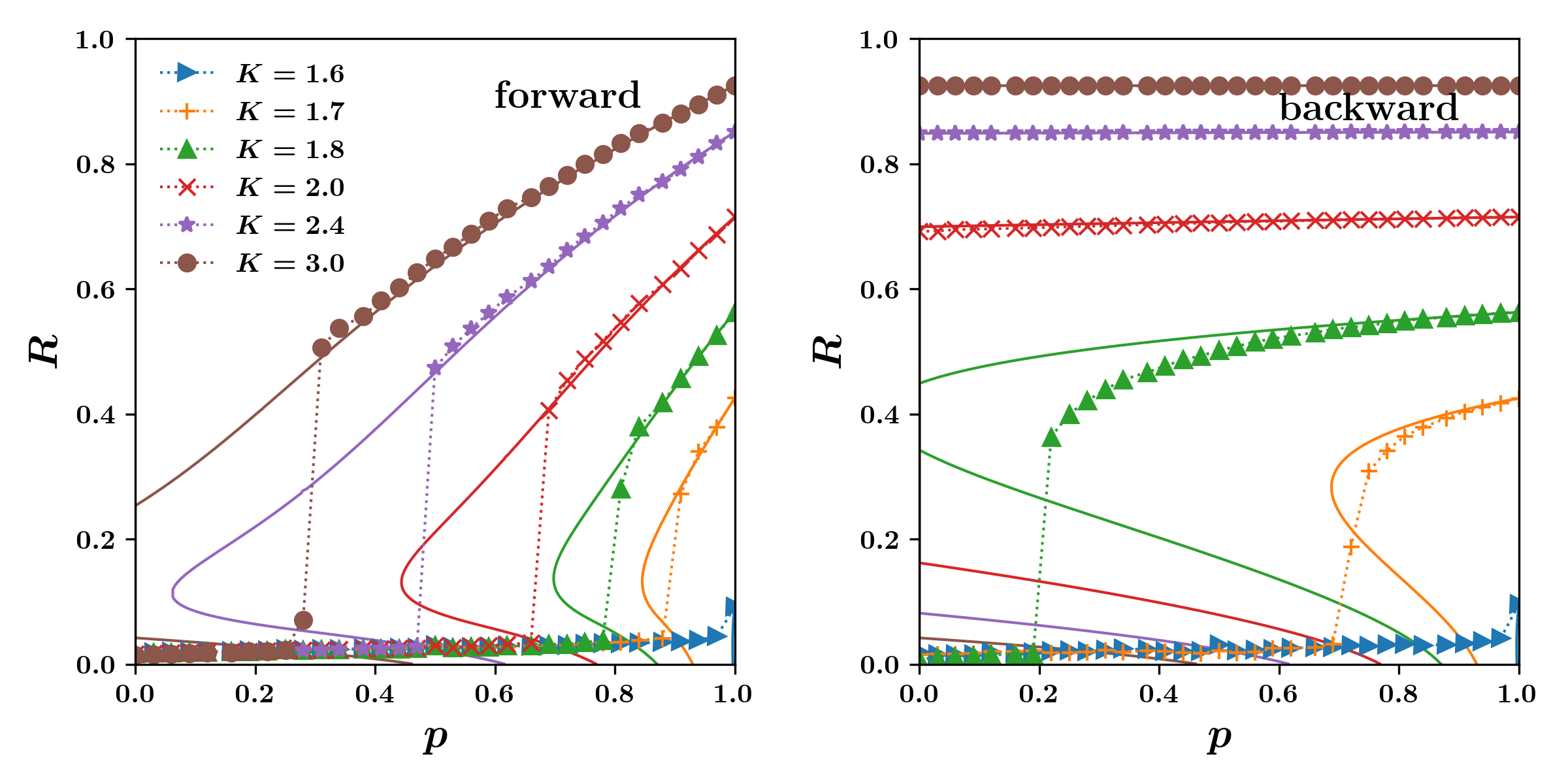}
\caption{
    The order parameter dependence on the fraction $p$ of renewable power source. Solid curves are solutions of Eq.~\eqref{eq:sce}, and symbols are finite-size ($N=1024$) numerical results. The coherence level remains higher in the backward process, and the forward process requires a larger coupling strength $K$ for the same level. Recall that the transition point of the first order KM ($p=1$) was $K_c=\sqrt{8/\pi}\approx1.596$. It is a minimal but insufficient strength $K$ for synchronizing a mixture system. Thus for $K=1.6$, supercritical regime, $p>p_c$ does appear, but it is quite limited near $p=1$. The first-order system is known to exhibit a continuous ST, and hysteresis is absent. Therefore, at $p=1$, the forward and backward values coincide.
}   \label{fig:p-vs-op}
\end{figure}

Next, we show that the coherence $R$ also depends on the fraction $p$, and this level of synchronization is path-dependent (see Fig.~\ref{fig:p-vs-op}), forward or backward.
Hysteresis is a path-dependent memory effect on the synchronization dynamics. Formation and disintegration of a synchronization cluster may unfold asymmetrically. In first-order equilibrium phase transitions, hysteresis is associated with latent heat. Here, the implications of the memory effect can be two sides; it can contribute to the resilience of a synchronous state upon a small perturbation, but it may also imply a cost when restored after a large-scale power outage. For the mixed-order model, the latent heat decreases; thus, the recovery becomes faster as $p$ increases.

A second-order or mixed-order Kuramoto equation, also called the swing equation from the electrical engineering context, describes the synchronization dynamics of a power grid~\cite{powergrid,powergrid2,rohden}. Generators with large turbines dominate traditional power grids and, thus, have large inertia. In contrast, the power produced from renewable sources such as solar photovoltaic panels and wind turbines is usually connected to the grid through electronic inverters with little rotational inertia. The modernized power grid puts increasingly more weight on renewable sources and keeps broadening the inertia spectrum. However, the physical characteristics of an oscillator or the whole oscillator system will be divided into overdamped or underdamped, and therefore, the current bipartisan mixed-order KM can capture the basic essence of the ST of the power grid.

Furthermore, it has been suggested that reducing total inertia in power grids may raise significant problems in maintaining stability~\cite{Dorfler,Laurent}. Our model faces similar repercussions. The grid that relies more on renewable sources also faces a larger day-to-night variability of effective renewable fraction $p(t)$, simply because they are not 24 hours operable. Unlike traditional grids, $\Delta p$ may grow large on the modernized grids. Fig.~\ref{fig:p-vs-op} suggests that large variations in $p$ may cause large variations in the synchronizability $R$. Also, the ramping dynamics of the coherence will be hysteretic. Depending on the day's $p_{max}$, the required lower bound of the coupling strength for maintaining a synchronized operation of a power grid can vary. The coupling strength between two buses $i$ and $j$ of the power grid is given as $K_{ij}=B_{ij}|V_i||V_j|$, where $V_i$ is the voltage at bus $i$ and $B_{ij}$ is the imaginary part of admittance of the transmission line $(i,j)$~\cite{Dorf}.
For instance, notice from Fig.~\ref{fig:p-vs-op}(a) that the required minimum voltage (corresponding to $K$) to maintain the coherence strength $R$ should be higher for a lower $p$. Thus, the voltage constrainments of the buses in the power grid need to be updated in accordance with the variation in $p_{max}$. Increase of renewable sources $p$ can enhance grid synchronizability on average, however, on the other hand, controllability or predictibility of the grid can become undermined due to increased variability of $p$.

Finally, cluster fragmentation~\cite{Olmi,Gao2} occurs in the extreme underdamped mixture systems, denoted by the double-star subscript. A staircase pattern in Fig.~\ref{fig:multiclusters}(a) reveals the coexistence of multiple giant frequency clusters. The order parameter of a multi-cluster system shows large temporal oscillations in the steady state~\cite{Tanaka1}. Single-cluster to multi-cluster phase transition occurs at $m_{**}(p)$ in Fig.~\ref{fig:critical-line} and $p_{**}(m)$ in Fig.~\ref{fig:multiclusters}(c). Such instability is existent if the system is extreme-underdamped, and solidarity is restored as the overdamped population $p$ increases [Fig.~\ref{fig:multiclusters}(b)]. The typical value of $m_{**}$ is greater than $5$, and it is deduced that the use of the self-consistency method is valid for low-inertia systems $m<m_{**}(p)$.

In summary, we investigate a mixed-order KM's synchronization transition (ST). The ST of the oscillator mixture is of various types. Continuous ST occurs only for the pure first-order system. We defined and calculated critical inertia, where the system forks into two distinct dynamical characters: the underdamped system has multistability in the coherent states, while the overdamped system does not. However, they both exhibit hysteresis in forward and backward driving. Predicting and controlling the state of an underdamped system is hardly manageable because it depends intricately on the initial states of the oscillators or the driving protocols. Nevertheless, in a backward protocol, where the coupling constant is adiabatically decreased from a supercritical ($K>K_c$) system, a hybrid critical exponent $\beta_R$ or $\beta_P$ is encountered in the synchronization or percolation order parameter curve. The precise value of the exponent can be calculated using the self-consistent mean-field theory. We found that the critical exponent $\beta$ varies continuously over $p$ in the case of the critically damped systems.

Our results are strongly coupled to the issues of green energy transition and grid modernization. A mix of energy sources with low- and high-carbon emissions introduces a broader spectrum of inertia of the grid's power generators. Without knowing detailed values of each inertia of turbines, we can still grasp how the phase transition type of the grid will turn out if we can find the critical inertia $m_*$ of the grid. Our result suggests that the change in low-inertia population $p$ introduces an abrupt change in the synchronizability of the power grid. The green energy transition, which aims for increased utilization of low-inertia power sources, can further increase the day-and-night variation in $p$. That is simply because solar panels cannot operate at night. The grid can become even more unpredictable if this effect combines with the path-dependence coherence of an underdamped system. Therefore, renewable power sources should be exploited cautiously to avoid stepping into the tipping point.

\section{methods}

\subsection{Memory Effect}
The memory effect of the system causes hysteresis. The precise path of the order parameter curve $R(K)$ depends not only on the initial condition $\{\theta_i,\dot\theta_i\}_{i=1}^N$ at $t=0$, but also on how $K$ is increased or decreased. Hence, we must fix our protocol on how to update $K$. We consider the following three types:
\begin{align}
     & K \to K + dK                                                                       & \textrm{forward driving protocol}      \\
     & K \to K - dK                                                                       & \textrm{backward driving protocol}     \\
     & K_0 \stackrel{+dK}{\longrightarrow} K_1 \stackrel{-dK}{\longrightarrow} K_2 \cdots & \textrm{protocol $K_0,K_1,K_2,\cdots$}
\end{align}
The forward and backward protocols are most commonly used, but one can use, for instance, the third class of protocol $K_0,K_1,K_2,\cdots$. In Fig.~\ref{fig:multistability}(b), we use a slightly different protocol, where the change of the direction is triggered instead by the coherence targets, i.e. $R_0=0,R_1,R_2,\cdots$.
When the system is underdamped, the hysteresis is evident from separating the self-consistent order parameter curves. The backward self-consistent order parameter curve $R_b(K)$ lies higher than the forward self-consistent order parameter curve $R_f(K)$. When the system is overdamped, the two curves $R_f(K)$ and $R_b(K)$ collapse into a single curve. The shape of the curve suggests a subcritical bifurcation. Indeed, a hysteresis occurs between forward and backward driving, and the phase transition is discontinuous overall. (see Fig.~\ref{fig:checking-sce-with-simulation}). However, we will show that the ST of an individual driving protocol may become hybrid.

\subsection{Self-consistency Solution}
\begin{figure}
\centering
\includegraphics[width=\linewidth]{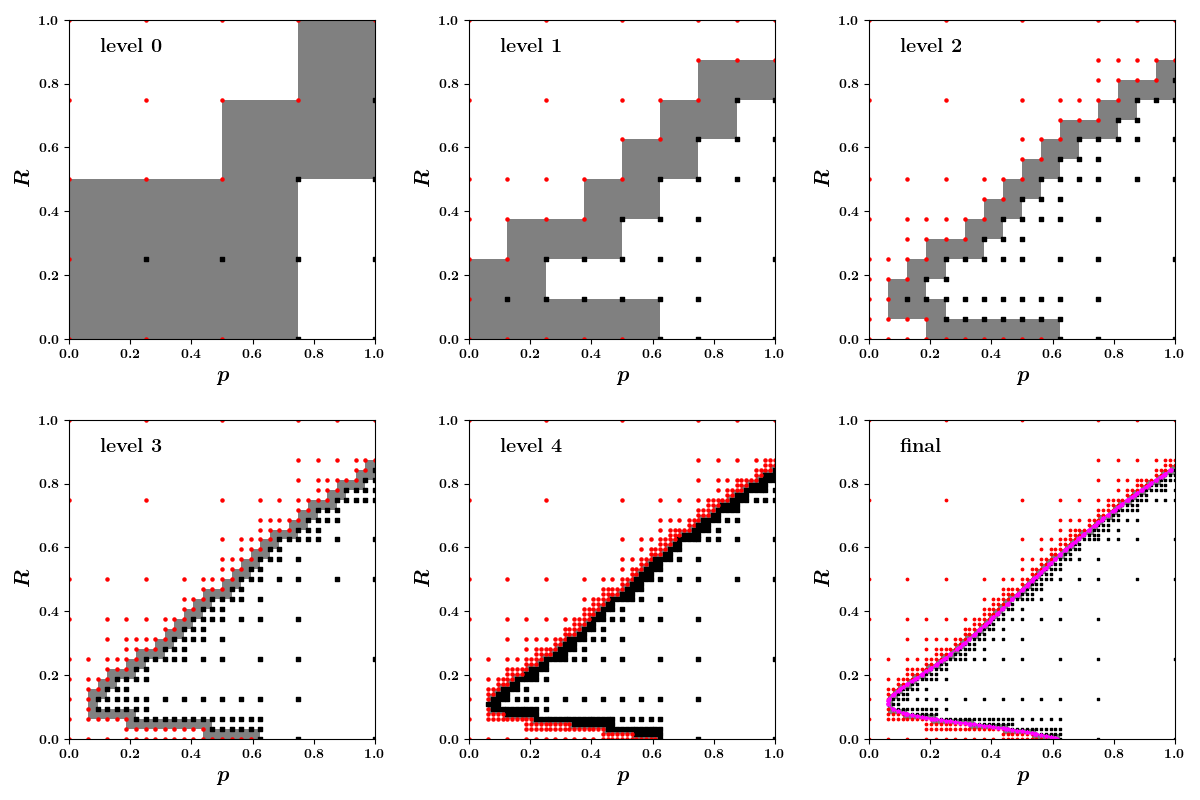}
\caption{
    An example demonstration of the divide and conquer algorithm to search solutions of the self-consistency equation on a two-dimensional parameter domain $(p,R)$. The iterative bisection can be stopped at a target resolution. In the final step, solutions are polished using any root-finding method that suits one's preference. Here, we used the secant method. This algorithm can be likewise extended to a multidimensional domain $(K,m,p,R)$.
}   \label{fig:sce-mesh-algorithm}
\end{figure}

The self-consistency order parameter solutions $R(K,m,p)$ form a hypersurface over the three-dimensional control parameter domain. The solution of Eq.~\eqref{eq:sce} is obtained using the bisection method on the line segments of a rectangular mesh in the space of $(K,m,p,R)$. A grid line contains a solution if the residuals $r(x)$ and $r(y)$ of Eq.~\eqref{eq:sce} at its two endpoints $x$ and $y$ have different signs $r(x)r(y)\leq 0$. For efficiency, we iteratively divide the mesh only at the locations containing solutions until the target resolution is reached. In the final step, the solutions are polished using the secant method. See Fig.~\ref{fig:sce-mesh-algorithm}. Having solved the self-consistent $R(K,m,p)$, we can use this value to determine the corresponding percolation order parameter $G(K,m,p)$ using Eq.~\eqref{eq:perc-op}.

\subsection{Numerical Integration}
The self-consistency solution is cross-checked with the steady-state long-time average value of the order parameter in the direct numerical integration of the finite-size model. Eq.~\eqref{eq:model} is solved using the fourth-order Runge-Kutta method with time step size $h=0.01$s and system size $N=1024$. We use regularly sampled values of $\omega_i$ for each population $N_1$ and $N_2$. $\gamma$ is set to unity. In the forward process, we initially choose random values of $\{\theta_i\}$ and zero phase velocities $\{\dot\theta_i=0\}$ at $K=0$. The coupling strength $K$ is increased by $\Delta K=0.1$, and the system is run for a time duration $T=10^3$s at each following step of $K$, and it is repeated until $K=15$ is reached. Afterward, the backward process is followed; $K$ is decreased back to $K=0$ by the amount $\Delta K=0.1$ at a step in a similar manner.

\subsection{Saddle}
To solve a saddle of $K(m,R)$, we introduce a little trick. We regard the surface $K(m,R)$ as a potential and consider the following relaxation dynamics, known as the gentlest ascent dynamics~\cite{GAD}.
\begin{align}   \label{eq:gad}
    \dot{x} & = -\nabla K(x) + 2 \frac{v^T \nabla K(x)}{v^Tv} v \nonumber \\
    \dot{v} & = - H(x) v + \frac{v^T H(x)v}{v^T v} v
\end{align}
where $x=(m,R)^T, v=\dot{x}$ and $H(x)$ is the Hessian matrix $\partial^2 K/\partial x_i \partial x_j$. Notice that the first term of the equation of $x$ corresponds to the usual gradient dynamics. The gradient descent method is used to obtain the local minimum of a basin. The second term introduces a slight modification to this dynamics. The role of the second term in the equation of $x$ is to give a correction in the updated direction so that the vector component of $x$ in parallel to $v$ is flipped to the direction opposite of $v$, while the other vector component of $x$ perpendicular to $v$ is left as is. In the equation of $v$, the first term subtracts $Hv$. The vector components of $v$ in parallel to eigenvectors of $H$ with larger eigenvalues will be subtracted more. Notice also that
\begin{align}
    \frac{d}{dt}\frac{\gamma}{2}v^T v = -v^T H(x) v + \frac{v^T H(x)v}{v^T v} v^T v = 0,
\end{align}
and therefore $|v|^2$ is a conserved quantity. Thus, the role of the second term in the equation for $v$ is also to maintain the normalization of $v$. The vector component of $v$, which corresponds to the eigenvector of $H$ with the smallest eigenvalue, is consequently reinforced throughout the relaxation dynamics~\eqref{eq:gad}. In the end, $v$ becomes the normalized Hessian eigenvector corresponding to the minimum eigenvalue. $x(t)$ follows the gradient ascent dynamics in the direction of negative $v$ and relaxes to the saddle $\nabla V(x)=0$.

\subsection{Measurements of $\beta_R$ and $\beta_G$}
We measure the critical exponent $\beta_R$ of the ST of the mixed-order KM from the self-consistent order parameter curves of the infinite system. Fig.~\ref{fig:measurement-of-beta} shows the order parameters $R$ when the second-order oscillators have [(a) and (d)] an overdamped inertia $m < m_*$ and [(b) and (e)] a critical inertia $m=m_*(p)$. The transition point $K_c$ and the jump size $R_c$ are measured by the location of the singular point $\partial K/\partial R=0$ on the order parameter curve $R(K)$, for each given $p$. When $m\leq m_*(p)$, the forward ($\omega_c=\omega_h$) and backward ($\omega_c=KR$) self-consistency solutions result in an identical order parameter curve. However, the numerical simulations reveal forward-backward hysteresis. The self-consistency solutions in the range $R\geq R_c$ ($0<R<R_c$) are assumed to be stable (unstable). The backward transition corresponds to a saddle-node bifurcation, giving a hybrid critical $\beta$ in the order parameter curve. In contrast, the forward transition is discontinuous with $\beta=1$ after the jump. See Supplementary Information for more details.

\section*{Acknowledgments}
B.K. was supported by the National Research Foundation of Korea by Grant No. RS-2023-00279802 and the KENTECH Research Grant No. KRG-2021-01-007.
The authors declare no conflicts of interest.


\clearpage
\newpage
\onecolumn
\begin{center}
    \textbf{\large Supplementary Information:\\
        Synchronization of Mixed-Order Kuramoto Oscillators in Decentralized Power Grid}
\end{center}
\setcounter{equation}{0}
\setcounter{figure}{0}
\setcounter{table}{0}
\setcounter{page}{1}
\makeatletter
\renewcommand{\thesection}{S\arabic{section}}
\renewcommand{\theequation}{S\arabic{equation}}
\renewcommand{\thefigure}{S\arabic{figure}}
\renewcommand{\thetable}{S\arabic{table}}
\renewcommand{\bibnumfmt}[1]{[S#1]}

In the supplementary information, we present i) measurement of critical exponents $\beta_R$ and $\beta_G$ in the mixed-order oscillator model and ii) frequency clustering of first- and second-order oscillators in the forward and backward processes for underdamped ($m=6)$ and the overdamped ($m=0.5$) cases.

\section*{Measurements of $\beta_R$ and $\beta_G$}
\begin{figure}[h!]
\centering
\includegraphics[width=\textwidth]{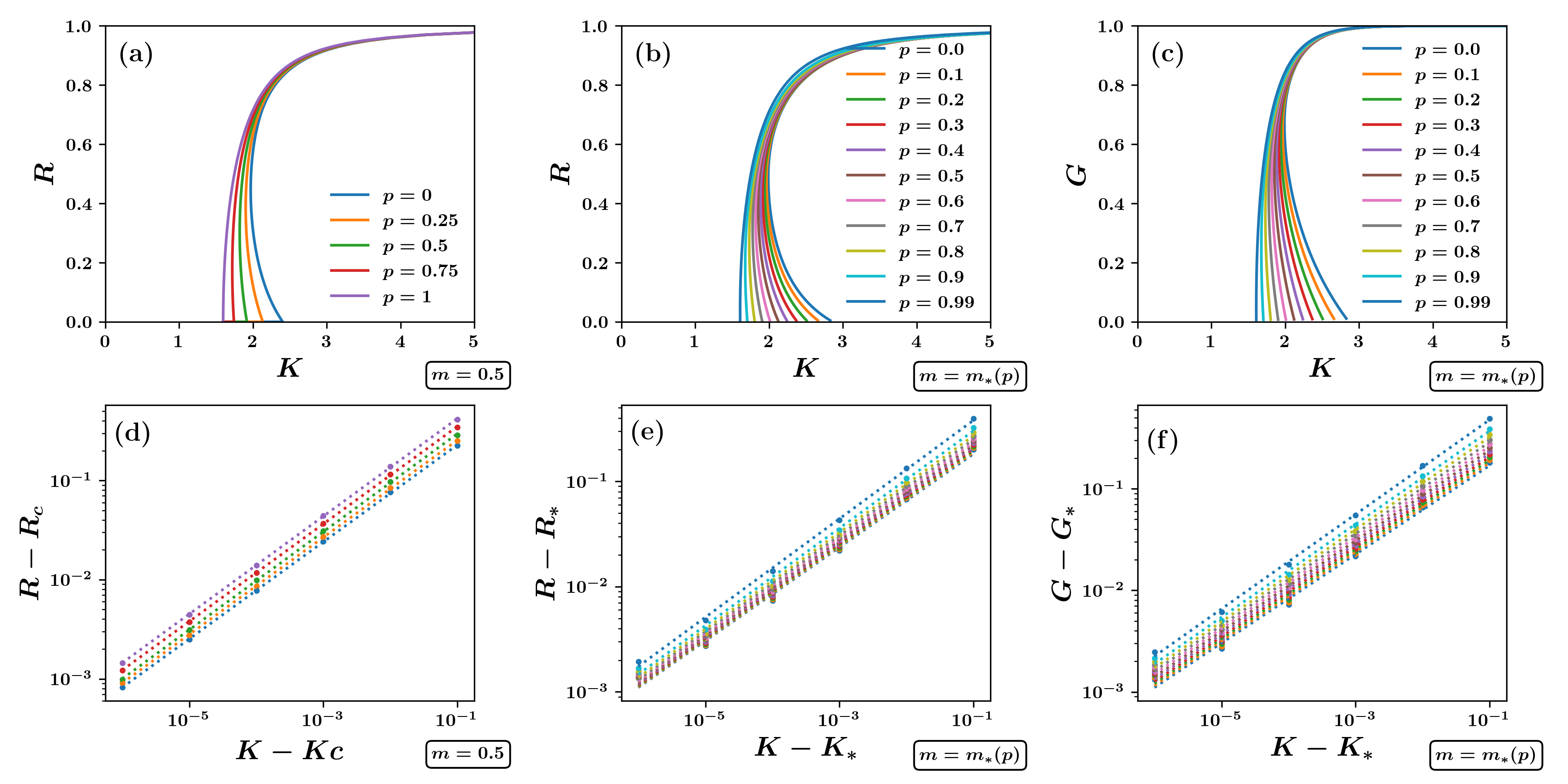}
\caption{
    Measurement of $\beta_R,\beta_G$: The order parameter curves obtained from the SCE~\eqref{eq:sce} for an overdamped inertia $m=0.5$ and critical inertia $m=m_*(p)$ vs the coupling strength $K$ are shown in (a) and (b), respectively. The transition point $K_c$ and jump size $R_c$ are defined by the location of the spinodal point $\partial K/\partial R=0$ for each order parameter curve for each order parameter curve $R(K;p)$. Following the definition Eq.~\eqref{eq:hpt}, the critical exponent $\beta_R$ is measured by the slope of the fitted line in the double logarithmic plots of $R-R_c$ versus $K-K_c$, as shown in (d) and (e). For the overdamped inertia $m=0.5$, the exponent $\beta_R$ is close to the continuous ST $\beta_R=1/2$ value. The exponent $\beta_R$ along the critical inertia curve ranges between $0.44$ and $0.5$. The percolation order parameter curve $G(K;p)$ calculated from Eq.~\eqref{eq:perc-op} and its critical exponent $\beta_G$ are shown in (c) and (f), respectively, for the critical inertia $m=m_*(p)$ at each $p$. (d) $\beta_R=0.49$ was obtained at $p=0,0.25,0.5,0.75,0.99$ (from bottom to top). (e) $\beta_R=0.44$ for $p=0$, $\beta_R=0.45$ for $p=0.1,0.2,0.3,0.4,0.5,0.6$, $\beta_R=0.46$ for $p=0.7,0.8,0.9$, and $\beta_R=0.47$ for $p=0.99$ (from bottom to top). (f) $\beta_G=0.44$ for $p=0,0.1,0.2,0.3,0.4$, $\beta_G=0.45$ for $p=0.5,0.6,0.7,0.8$, $\beta_G=0.46$ for $p=0.9$, and $\beta_G=0.47$ for $p=0.99$ (from bottom to top).
}   \label{fig:measurement-of-beta}
\end{figure}

We measure the critical exponent $\beta_R$ of the ST of the mixed-order KM from the self-consistent order parameter curves of the infinite system. Fig.~\ref{fig:measurement-of-beta} shows the order parameters $R$ vs $K$ when the second-order oscillators have [(a) and (c)] an overdamped inertia $m < m_*$ and [(b) and (d)] a critical inertia $m=m_*(p)$. The transition point $K_c$ and the jump size $R_c$ are measured as the location of the singular point $\partial K/\partial R=0$ on the order parameter curve $R(K)$, for each given $p$. When $m\leq m_*(p)$, the forward self-consistency order parameter curve ($\omega_c=\omega_h$) and backward self-consistency order parameter curve ($\omega_c=KR$) are identical. However, hysteresis is present. The self-consistency solutions in the range $R\geq R_c$ ($0<R<R_c$) are assumed stable (unstable). It is remarked, however, that hysteresis may appear in finite systems due to finite size and finite time effects.

\section*{Critical inertia curve}
\begin{figure}[h!]
\centering
\includegraphics[width=0.5\textwidth]{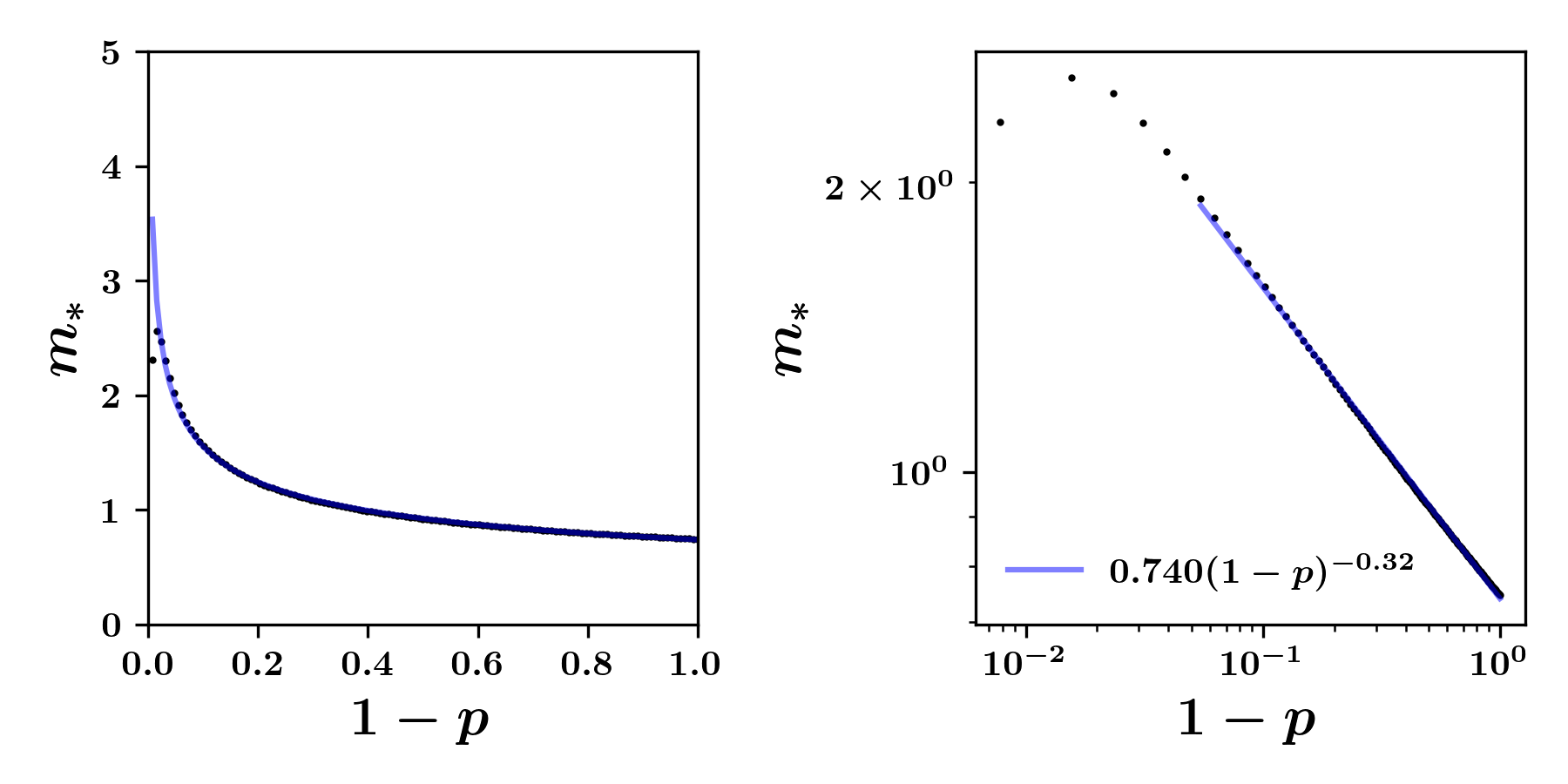}
\caption{
    The plot of $1-p$ versus $m_*$ and its double logarithmic plot. The critical inertia of the mixed-order KM fits to the simple algebraic from $m_*(p) \approx 0.740 (1-p)^{-0.32}$ except for when $p$ is very close to unity, which corresponds to the first-order KM.
}   \label{fig:mstar-fitted}
\end{figure}

We notice from Fig.~\ref{fig:mstar-fitted} that the curve of critical inertia $m_*(p)$ is fitted as the following form
\begin{align}
    m_*(p) \approx 0.740 (1-p)^{-0.32}.
\end{align}
When $1-p$ is increased, the fraction of underdamping oscillators increases in the mixture system. In turn, the critical inertia of the second-order oscillators is decreased. It is noticed that the exponent is not $-1$. Hence, the region of overdamping or underdamping cannot be distinguished using a single $p$-independent characteristic scalar value $\langle m \rangle$. Instead, the average characteristic mass of the system also scales as $(1-p)m_* \propto (1-p)^{0.68}$ with increasing fraction of the inertial oscillators $1-p$.

\begin{figure}[h!]
\centering
\includegraphics[width=0.75\textwidth]{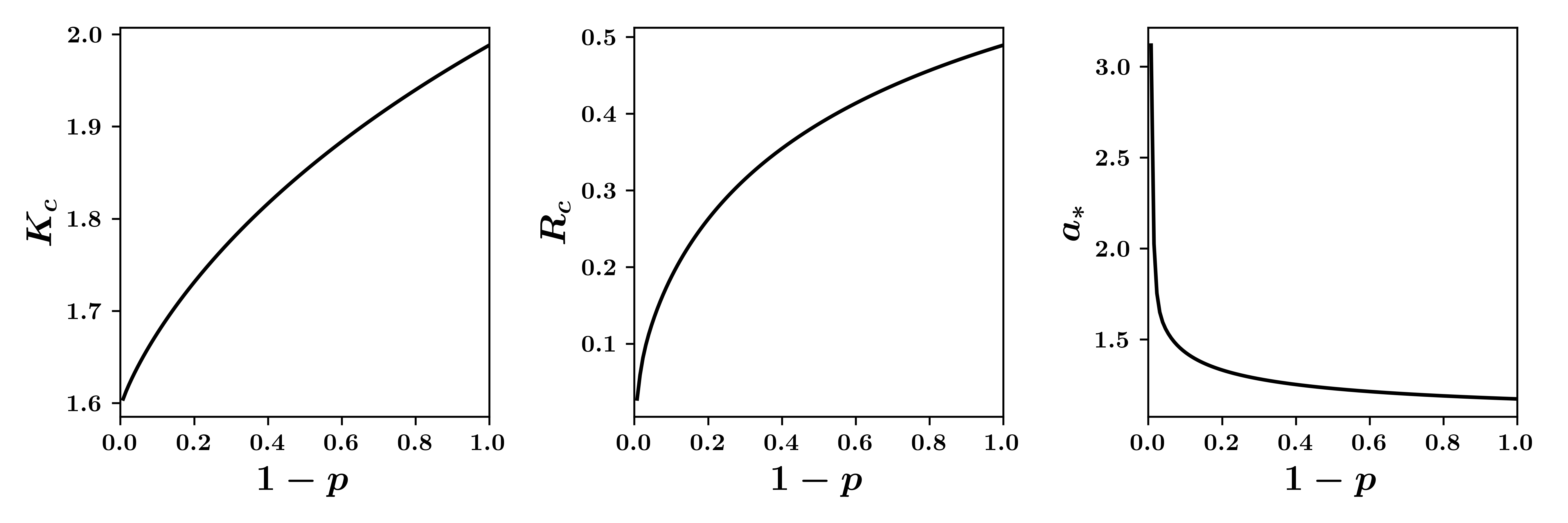}
\caption{
    Dependence of the transition point $K_c$, $R_c$ and $a_*=\gamma/\sqrt{m_*K_cR_c}$ versus $1-p$, along the critical inertia $m_*(p)$.
}   \label{fig:critical-points}
\end{figure}

\clearpage
\section*{Frequency-based clustering of oscillators}
ST occurs for an overdamped inertia $m=0.5<m_*$. It is remarked that $m_*\approx 0.92$ for the $p=0.5$ system. Notice from Fig.~ref{fig:clustering-overdamped} that the forward and backward locking and unlocking occur at the same transition point $K_c$, and the clustering patterns are almost identical for both first- and second-order oscillators. In the overdamped region, the forward self-consistency order parameter curve is identical to the backward one. The overdamped region $m<m_*$ corresponds to $a>a_*$ region of the single oscillator reduced phase (bifurcation) diagram of the second-order oscillators, where $a_*$ is the location where homoclinic, saddle-node, and infinite-period bifurcations all meet together. When $m<m_*$ or $a>a_*$, the multistability of the mixture system or the bistability of an individual second-order oscillator is absent.

\begin{figure}[b!]
\centering
\includegraphics[width=0.5\textwidth]{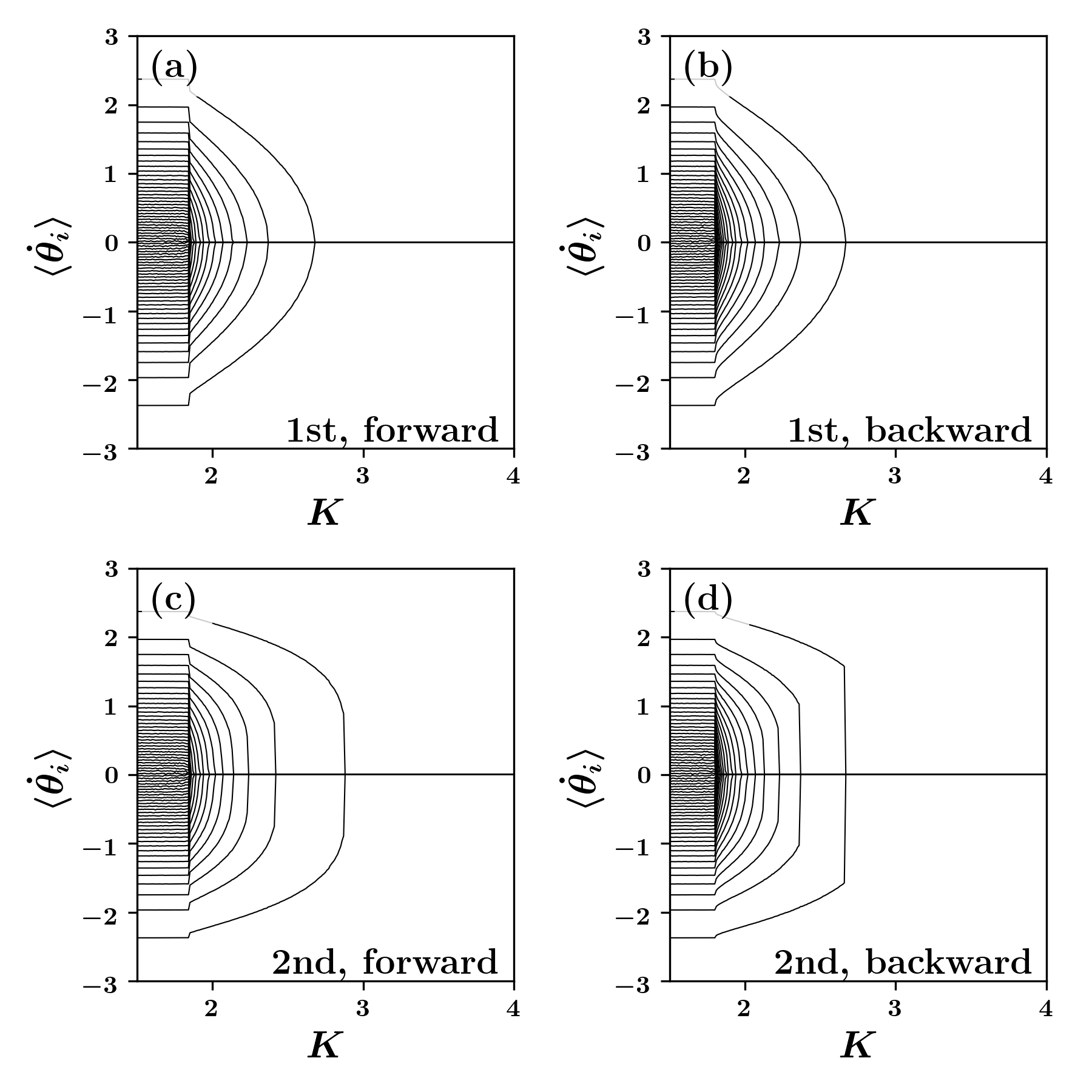}
\caption{
    Frequency clustering of oscillators when $m=0.5$ and $p=0.5$: The critical inertia of the $p=0.5$ system is about $m_*\approx 0.92$ and $m=0.5$ case constitutes an overdamped system. It is noticed that the forward locking and backward unlocking occur at nearly the same transition point, and the clustering patterns are almost identical for both first- and second-order oscillators. Therefore, the hysteresis in the order parameter curves becomes very small or almost absent. However, notice a small difference in the paths of some second order oscillators at the boundary. The backward velocity locking to $\langle\theta_i\rangle=0$ is sustained for a longer range of $K$ until a sudden jump occurs, which means a detachment of an oscillator from the giant cluster. This oscillator carries a memory of its locking-drifting status. $N_1=512, N_2=512$. Plotted every 8 oscillators for visual clarity.
}   \label{fig:clustering-overdamped}
\end{figure}

Bistability is absent for an overdamped mixed-order Kuramoto system in the region $m<m_*$, and the forward and backward clustering patterns are similar. See Fig.~\ref{fig:clustering-overdamped}(c) and (d). There, a small difference is noticed in the forward and backward clustering paths of a second-order oscillator with a large natural frequency $|\omega_i|$, and it is thought to be a finite-size effect.

\begin{figure}[t!]
\centering
\includegraphics[width=0.8\textwidth]{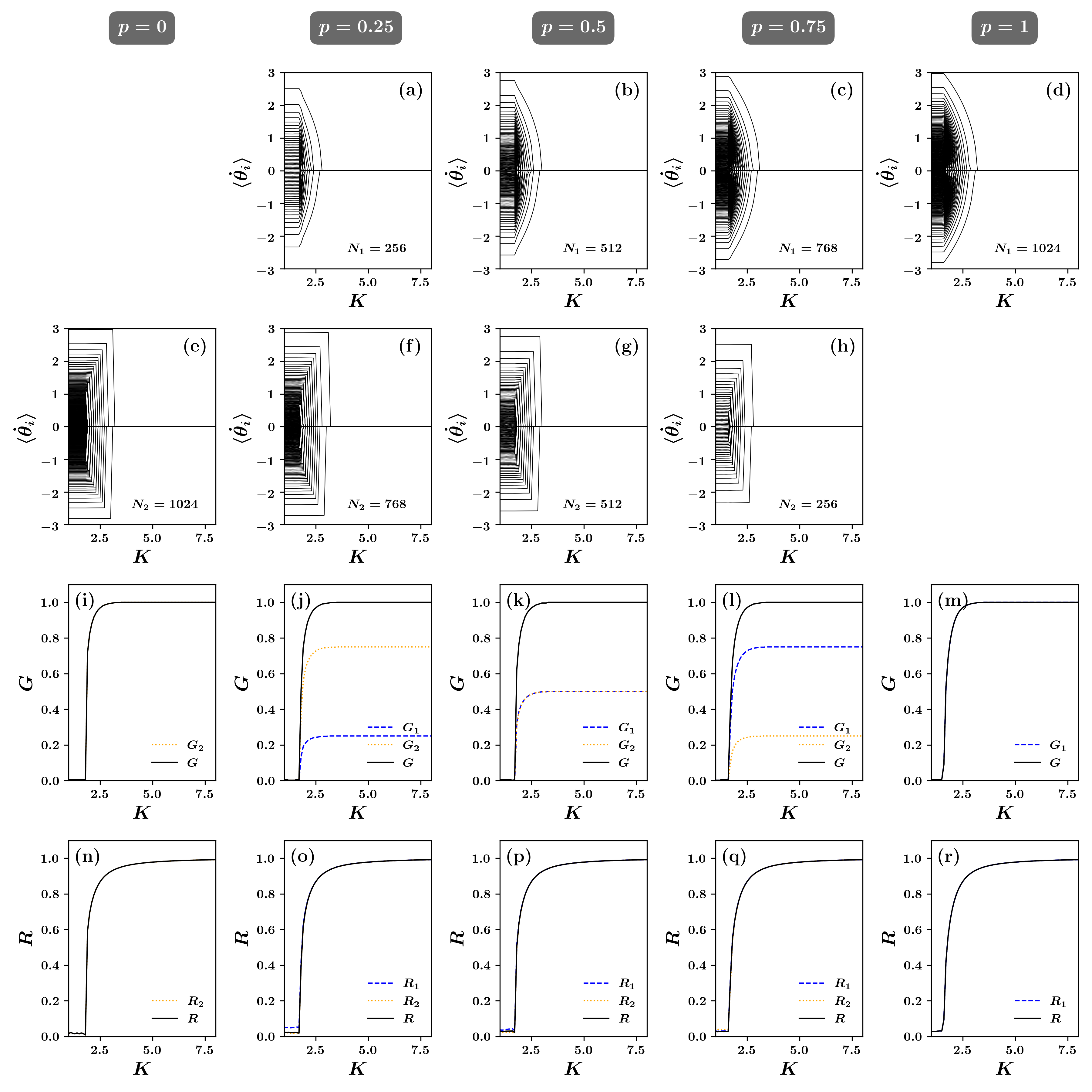}
\caption{
    Frequency clustering of oscillators in the backward process ($m=6$): Notice, in comparison with Fig.~\ref{fig:clustering-underdamped-forward}, the hysteresis in drop and jump of coherence. $K_c^b<K_c^f$. The order parameter curves of first- and second-order oscillators $R_1$, $R_2$ in the backward process are almost identical, and thus $R_1 \approx R_2 \approx R$. The hysteresis between the forward process (Fig.~\ref{fig:clustering-underdamped-forward}) and the backward process shows that the mixture system becomes underdamping when $m=6$. $m=6$ corresponds to a discontinuous ST region in the Fig.~\ref{fig:critical-line}.
}   \label{fig:clustering-underdamped-backward}
\end{figure}

Each column of Fig.~\ref{fig:clustering-underdamped-backward} corresponds to the simulation result of finite $N=1024$ systems, ranging from a pure second-order ($p=0$), mixed-orders $25:75$, $50:50$, $75:25$, and a pure first-order ($p=1$). The first and second rows show the microscopic frequency clustering patterns of first- and second-order oscillators. The third row corresponds to the percolation order parameter, which is defined as the fraction of oscillators included by the giant cluster ($G$) and a fraction of first-order/second-order oscillators included by the giant cluster ($G_1$/$G_2$). It is mentioned that $ N$ normalizes the number of oscillators for the percolation order parameters, thus $G=G_1+G_2$. The fourth row is the Kuramoto order parameter $R=(1/N)|\sum e^{i\theta}|$, and $R_a=(1/N_a)|\sum_a e^{i\theta}|$ where summation with subscript $a=1,2$ applies to first- and second-order oscillators respectively. It is remarked that the normalization is defined differently ($N, N_1, N_2$) for the synchronization order parameters. The three synchronization order parameter curves coincide in the backward process $R(K)$. This is because the saddle-node bifurcation line $\omega_c = KR$ is followed for the second-order oscillators in the backward process, which is identical to the boundary of the locking condition $\omega_c \leq KR$ of the first-order oscillators. In other words, the steady states of all of the bistable second-order oscillators, which may either become fixed at a constant $\theta$ or follow a limit cycle motion $\dot\theta(\theta)$ in the steady state, have been set to fixed states.

\begin{figure}[t!]
\centering
\includegraphics[width=\textwidth]{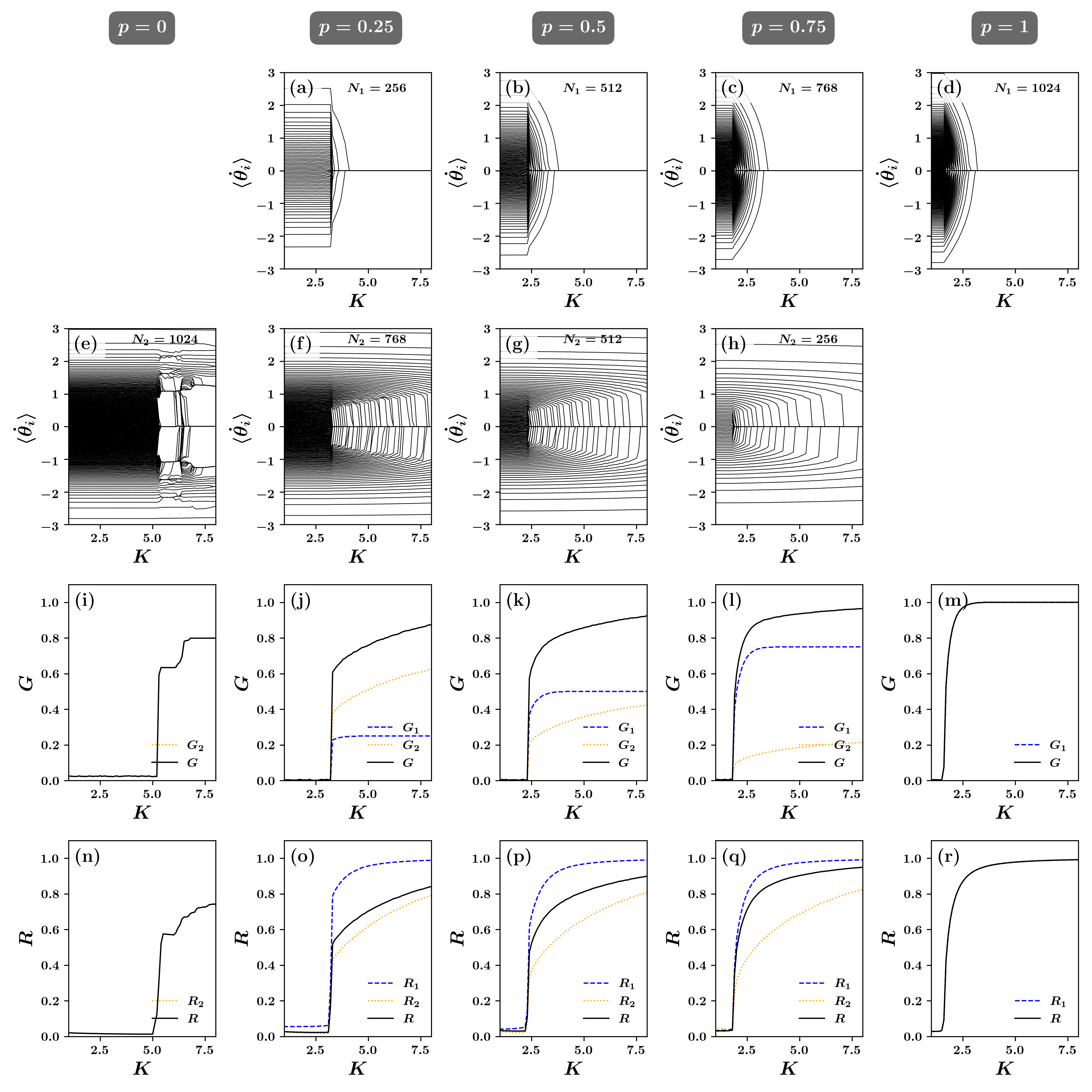}
\caption{
    Frequency clustering of oscillators in the forward process ($m=6$): Time averaged angular velocities of (a)--(d) first- and (e)--(h) second-order oscillators $\langle\dot\theta_i\rangle$ versus $K$. For visual clarity, oscillators are sparsely sampled (every $8$). (i)--(m) fraction of frequency locked oscillators (percolation order parameter) $G,G_1,G_2$ and (n)--(r) the synchronization order parameters $R,R_1,R_2$. Columns correspond to $p=0,0.25,0.5,0.75,1$. (e) At the transition point of the $p=0$ system, the primary frequency cluster emerges at $\langle \dot\theta\rangle=0$, and also secondary and higher-order clusters emerge. As $K$ increases beyond the transition point, secondary and higher-order clusters eventually disappear and merge with the primary cluster. Consequentially, a staircase pattern in Fig.~\ref{fig:multiclusters} is generated. (f)--(h) For $p>0$ systems, secondary clusters are not noticed. Total number of oscillators is $N=1024$. $\gamma=1, m=6$.
}   \label{fig:clustering-underdamped-forward}
\end{figure}

\begin{figure}
\centering
\includegraphics[width=\textwidth]{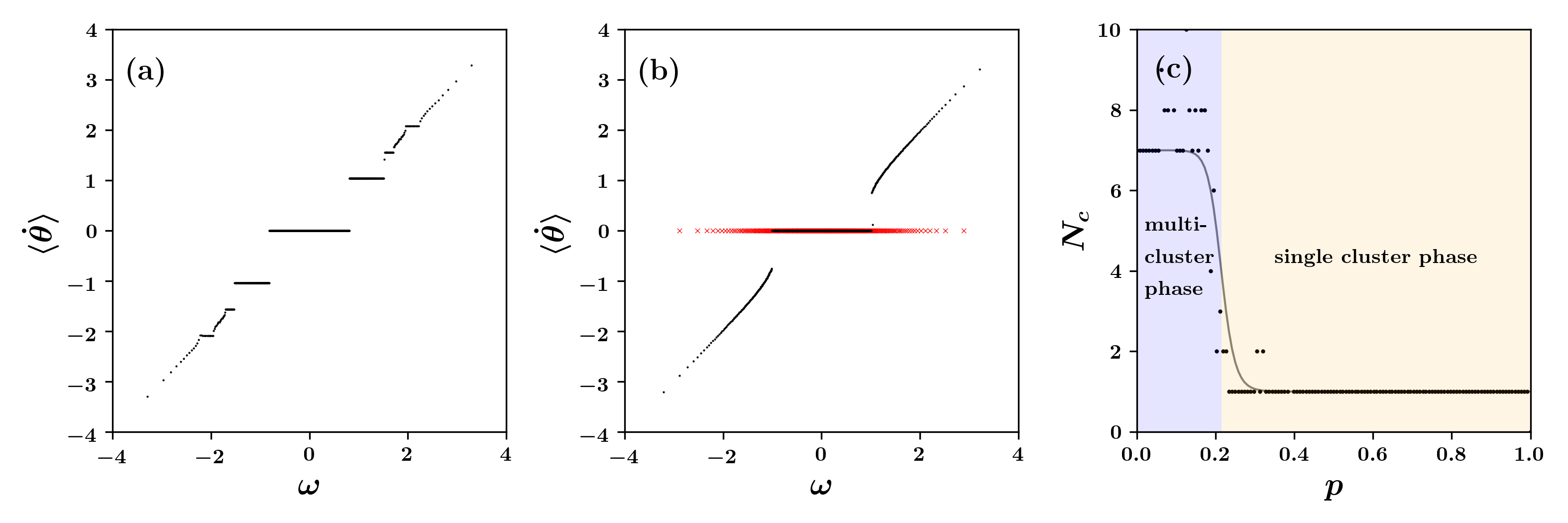}
\caption{
    Multiple giant frequency clusters can coexist in the mixture model for strongly underdamped ($m>m_{**}$) case. (a) Time-averaged angular frequency of individual oscillators $\langle \dot\theta\rangle$ versus $\omega$ reveals a staircase pattern at $p=0, m=6, K=5$. Multiple frequency clusters coexist in the coherent phase after the synchronization transition. (b) A single synchronized cluster of zero average frequency $\langle \dot\theta\rangle = 0$ exists in the coherent phase after synchronization transition for $p=0.25, m=6$ at $K=5$. The black dots (red crosses) correspond to the second-order (first-order ) oscillators. (c) Number of frequency clusters $N_c$ versus $p$ for $m=6$. To obtain the fitted curve, $N_c$ is clipped and normalized to $n_c$ in the range $[\epsilon,1-\epsilon]$ with a sufficiently small $\epsilon=1\mathrm{e}{-5}$, and then the corresponding logit function $n_c(p)/(1+n_c(p))$ was regressed to a quadratic function $a_0 + a_1 p + a_2 p$. The fitted curve is parameterized as $n_c(p) = 1/(1+\exp[-(a_0 + a_1 p + a_2 p^2)])$, with $a_0 = 15.281, a_1=-84.346, a_2=61.262$. The result indicates that $p_{**}(m=6)\approx 0.215$ and $m_{**}(p=0.215)\approx 6$.
}   \label{fig:multiclusters}
\end{figure}

\begin{figure}[t!]
\centering
\includegraphics[width=0.5\textwidth]{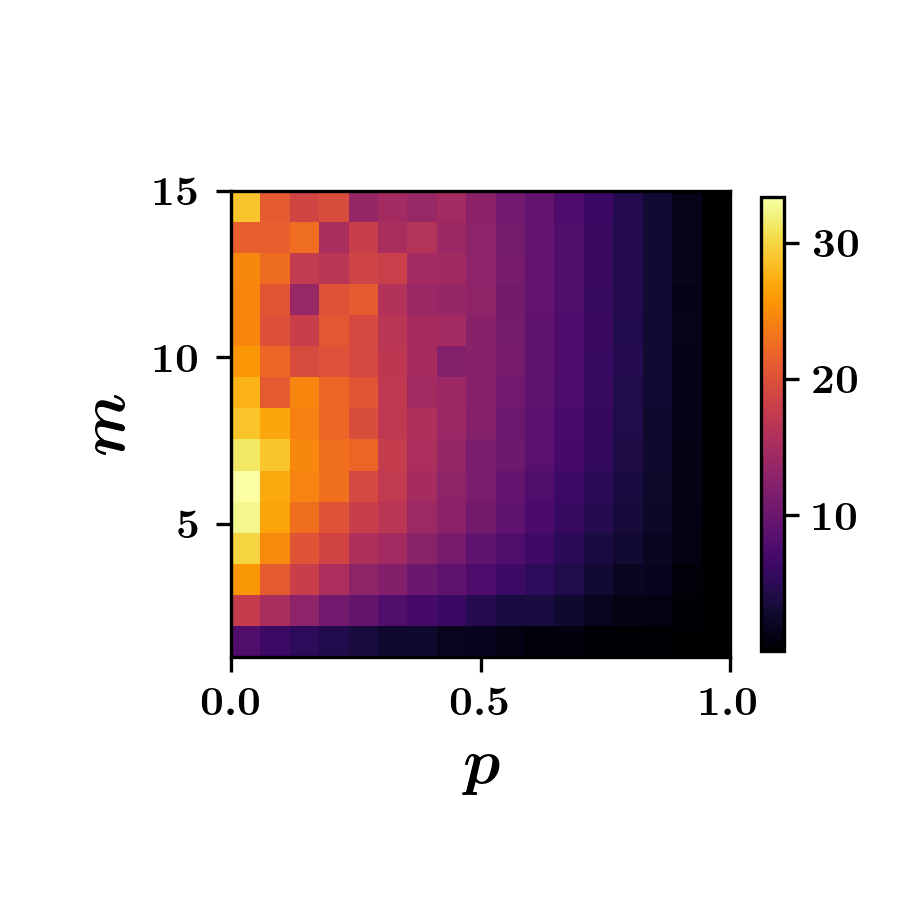}
\caption{
    Measurement of the size of hysteresis between forward and backward order parameter curves obtained from the Runge-Kutta simulation at each $(p,m)$ for $K \in [0,10]$. $N=1024$. Notice that the size of hysteresis decreases near the pure first-order model, at $m=0$ or $p=1$, which exhibits a continuous ST.
}   \label{fig:hysteresis}
\end{figure}

In contrast to a backward process, a nonstationary steady state with temporally oscillating order parameter $R(t)$ may appear in the forward process for a sufficiently large inertia $m>m_{**}$. Indeed, when $p=0$, one can notice at least three frequency clusters of the second-order oscillators after the ST [Fig.~\ref{fig:clustering-underdamped-forward}(e)]; one is at the zero level $\langle\dot\theta\rangle=0$ and the others are at the level $\pm 1.1$ at $K=5$. The coexistence of multiple giant frequency clusters in the system is revealed more clearly from the staircase pattern shown in Fig.~\ref{fig:multiclusters}(a). We notice secondary and higher-order clusters are frequency-locked at multiple levels of frequencies. The staircases can be calculated using Melnikov's method. Large-size synchronization clusters rotating at different frequencies generates a large-size oscillation of the order parameter $R(t)$ in the steady state. Thus, the system is synchronized to a non-stationary steady state after $K_c$. In contrast, for $p=0.25$ and $m=6$ in Fig.~\ref{fig:clustering-underdamped-forward}(f), only a single frequency cluster is found after the synchronization transition, as noticed in Fig.~\ref{fig:multiclusters}(b). A transition occurs from a multiple frequency cluster steady state to a single cluster steady state, as shown in Fig.~\ref{fig:multiclusters}(c). This transition between a non-stationary coherent phase to a stationary coherent phase is denoted by the double stars subscript and parameterized as $p_{**}(m)$ or $m_{**}(p)$. For $m=6$ we note that the multi-cluster to single cluster transition occurs at $p_{**}\approx 0.215$. Therefore, for mixture systems with $p=0.25,0.5,0.75$ and $m=6$, in Fig.~\ref{fig:clustering-underdamped-forward}, the clusters gather into a single line $\langle \dot\theta\rangle$, and secondary and higher-order frequency clusters are not recognized. In panel (a), the first-order oscillators undergo a sudden phase locking at the transition point. For the $p=1$ continuous transition, panel (d) shows the continuous growth of the frequency cluster after the transition.

$G$ in the third row is the percolation order parameter of the giant frequency cluster. It is defined as the fraction of oscillators that belongs to the largest frequency cluster, which rotates at angular velocity $\langle\dot\theta\rangle=0$. $G$ is further decomposed into first- and second-order populations $G_1$ and $G_2$, which correspond to the first and second term of Eq.~\eqref{eq:perc-op}. Hence $G=G_1+G_2$. After the ST, the size of the cluster $G$ continues to grow by eating the remaining nodes in a fashion similar to a continuous synchronization phase transition. As coupling strength $K$ increases, all oscillators will eventually attach to the giant cluster. However, the attachment of second-order oscillators occurs much more slowly than the attachment of first-order oscillators. First-order oscillators phase lock more quickly. The order parameters $R,R_1,$ and $R_2$ will continue to grow towards unity, even after $G$ has reached unity through a full phase-locked state; however because the spread of the phases $\theta_i$ can continue to decrease until it reaches the full phase synchrony.

The order parameters $R, R_1, R_2$ are shown in the last row. It is remarked that the emergence of coherence of first-order oscillators $R_1$ and the emergence of coherence of the second-order oscillators $R_2$ occur together at the same transition point. Therefore, it is sufficient to describe the phase transition of the mixed-order oscillator system using only a single order parameter $R$ instead of both $R_{1,2}$. As noticed from the supercritical regimes of the panels (a)--(h), the remaining drifting first-order oscillators cluster fast, while the locking of remaining drifting second-order oscillators is lagged. Oscillators with inertia cluster more gradually, so $R_1$ is larger than $R_2$, and the average $R=pR_1+(1-p)R_2$ lies in between. 

\clearpage
\section*{Instability}
\begin{figure}[h!]
\centering
\includegraphics[width=0.8\textwidth]{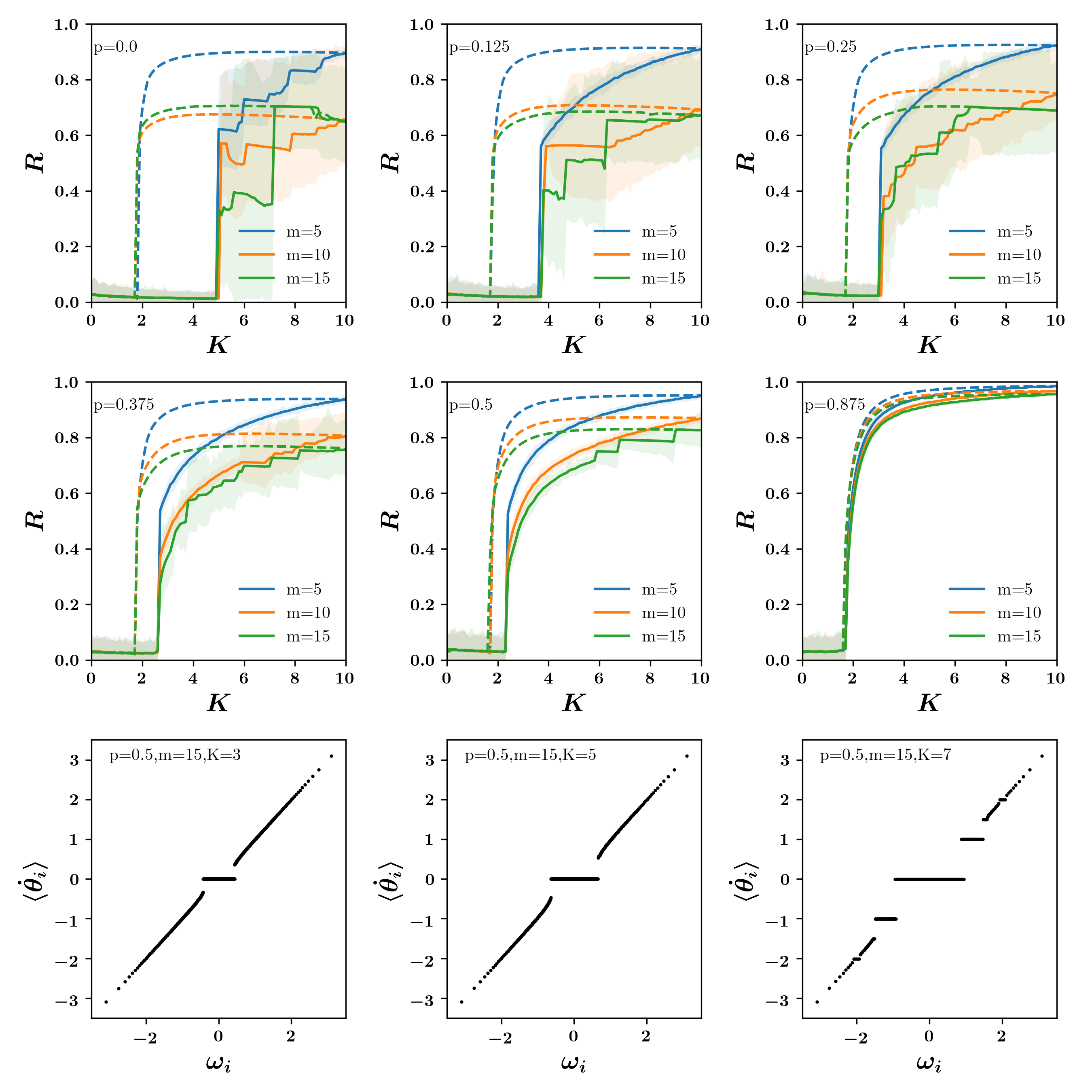}
\caption{
    (a)--(f) corresponds to the forward and backward order parameters curves $R(K)$ for different values of inertia $m=5,10,15$. For smaller $p$ and higher $m$, the order parameters can show large-scale oscillations in the steady state. For (a)--(d), the phase of the mixture system changes from IC$\to$MC. Multiple frequency clusters will eventually merge into one if $K$ is further increased beyond $10$. We may obtain a transition MC$\to$SC at a much higher $K$. The case of (e), especially when $p=0.5$ and $m=15$, is quite different. We observe consecutive transitions IC$\to$SC$\to$MC. Here, a single cluster state emerges at the synchronization transition to the coherent phase at $K_c$. However, there seems to be instability, and the single cluster state undergoes another transition to a multi-cluster state at a higher $K$. In (g)--(i), we plot the staircase pattern at $K=3,5,7$ for $p=0.5$. In (g) and (f), the width of the locked clusters grows after $K_c$. In (i), a multi-cluster pattern emerges.
}   \label{fig:instability}
\end{figure}

Finally, it is mentioned that the synchronization phase transition at $K_c$ in the region above $m>m_{**}$ transition curve occurs in two different patterns. In one case, the steady state of the mixture system progresses from an incoherent (IC) state to a multiple frequency cluster coherent state (MC) to a single frequency cluster coherent state (SC) as the coupling strength $K$ is increased. In another case, the progression is from IC to SC to MC. In the former case, MC emerges at $K_c$, while in the latter case, SC emerges at $K_c$, and MC emerges at some $K>K_c$. See Fig.~\ref{fig:instability}.

\clearpage
\section*{More order parameter curves}
A continuous phase transition occurs if $p=1$ (or $m=0$). It corresponds to the first-order KM. There, $\beta=1/2$ and $K_c=\sqrt{8/\pi}\approx 1.596$. The steady state of the infinite size system is stationary, i.e. the order parameter is constant in time $R(t)=R$. For $K<K_c$, the incoherent state (IC) with $R=0$ is the steady state. For $K>K_c$, coherent state(C) with $R>0$ is the steady state.

Otherwise, the mixed-order KM exhibits hysteresis in the forward and backward processes. Thus, a discontinuous phase transition occurs. In the backward process, the order parameter curve has a singularity with a critical exponent $\beta$ just before a discontinuous drop to $R=0$. $\beta$ is continuously varying in the range $[0.44,0.47]$ along the critical mass curve $m=m_*(p)$, otherwise $\beta$ is $1/2$.
\begin{figure}[h!]
\centering
\includegraphics[width=0.6\textwidth]{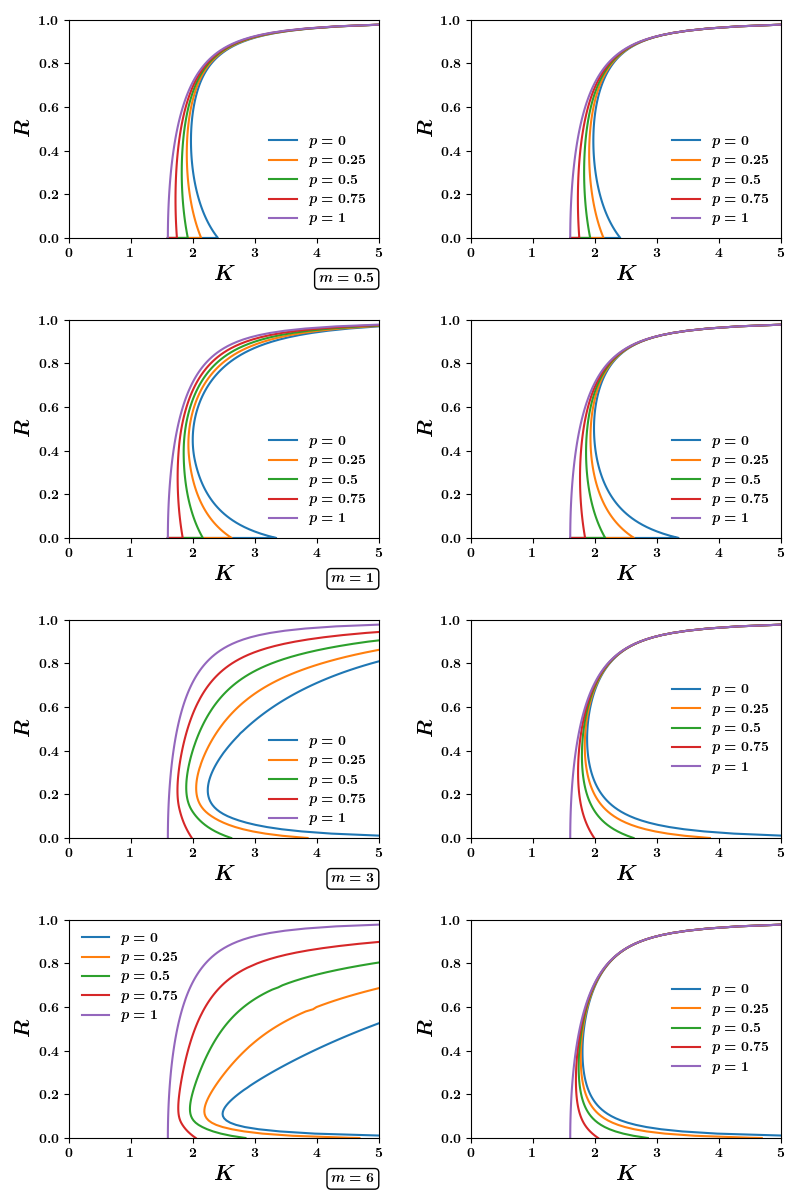}
\caption{
    Forward and backward self-consistent order parameter curves when $m=0.5,1,3,6$ and $p=0,0.25,0.5,0.75,1$.
}   \label{fig:more-curves}
\end{figure}

\end{document}